\documentclass[12pt,draftcls,journal,letterpaper,twosides,onecolumn]{IEEEtran}

\usepackage{amsmath,amssymb,amscd,latexsym,dsfont,wasysym,bbm}
\usepackage{float}
\usepackage{color}
\usepackage{graphicx,subfigure}
\usepackage{multirow,multicol} 
\usepackage{verbatim}
\usepackage{psfrag}
\usepackage{comment}
\usepackage{makeidx}
\usepackage[]{algorithm}
\usepackage{algorithmicx}
\usepackage{algpseudocode}
\usepackage{cite}
\usepackage{cases}
 \usepackage{setspace}
\usepackage{pgfplots} 

\newcommand{\tr}[1]{\textrm{#1}}
\newcommand{\mr}[1]{\mathrm{#1}}
\newcommand{\tnr}[1]{{\textnormal{#1}}}

\newcommand{\mc}[1]{\mathcal{#1}}
\newcommand{\mf}[1]{\mathsf{#1}}
 
\newcommand{\ms}[1]{\mathds{#1}}

\newcommand{\bc}{\boldsymbol{c}}

\newcommand{\bx}{\boldsymbol{x}}

\newcommand{\by}{\boldsymbol{y}}

\newcommand{\figref}[1]{Fig.~\ref{#1}}

\newcommand{\secref}[1]{Sec.~\ref{#1}}
\newcommand{\appref}[1]{Appendix~\ref{#1}}

\newcommand{\jj}{\jmath}      

\newcommand{\ie}{i.e.,~} 		
\newcommand{\eg}{e.g.,~}	

\newcommand{\convop}{\mathop{*}} 
\newcommand{\argmax}{\mathop{\mr{argmax}}}
\newcommand{\argmin}{\mathop{\mr{argmin}}}

\newcommand{\set}[1]{\{#1\}}
\newcommand{\SET}[1]{\left\{#1\right\}}

\newcommand{\cd}{\cdot}
\newcommand{\ld}{\ldots}

\newcommand{\e}{\mr{e}}

\newcommand{\PR}[1]{\Pr\SET{#1}}       	
\newcommand{\pdf}{f}            			

\newcommand{\IND}[1]{\ms{I}\big[{#1}\big]}   	
\newcommand{\Ex}{\ms{E}}     			
\newcommand{\dd}{\,\mr{d}}             		

\newcommand{\mcA}{\mc{A}}

\newcommand{\mcF}{\mc{F}}

\newcommand{\mcT}{\mc{T}}

\newcommand{\mfm}{\mf{m}}

\newcommand{\mfv}{\mf{v}}

\newcommand{\mfM}{\mf{M}}

\newcommand{\Natural}{\mathbb{N}}		
\newcommand{\SNRrv}{\mathsf{SNR}}  

\newcommand{\PER}{\mr{PER}}  

\newcommand{\No}{N_{0}}             
\newcommand{\Nb}{{\mathop{N_\tnr{b}}}} 	
\newcommand{\Ns}{{\mathop{N_\tnr{s}}}} 	

\newcommand{\Nc}{N_\tnr{c}} 	

\newcommand{\MOD}[1]{\Phi[#1]}      
\newcommand{\MODwo}{\Phi}      		
  
\pgfplotsset{compat=1.12}
\usetikzlibrary{positioning,shadows,shapes,fit,arrows,backgrounds}

\tikzstyle{rect_my} = [draw, rectangle, minimum width=2cm, text width=1.8cm, fill=gray!15, 
  text centered,  minimum height=.9cm]
\tikzstyle{square_my} = [draw, rectangle, minimum width=1cm, text width=0.8cm, fill=gray!15, 
  text centered,  minimum height=.9cm]
\tikzstyle{square_my_graph} = [draw, rectangle, minimum width=1.2cm, text width=1cm, fill=gray!15, 
  text centered,  minimum height=1.2cm]
\tikzstyle{circle_my} = [draw, circle, minimum width=1cm, text width=0.8cm, fill=gray!15, 
  text centered,  minimum height=.9cm]
\tikzstyle{circle_my_graph} = [draw, circle, minimum width=1.1cm, text width=.8cm, fill=gray!15, 
  text centered]
\tikzstyle{cloud_my} = [draw, shape=cloud, minimum width=1cm, text width=0.8cm, fill=gray!15, 
  text centered,  minimum height=1cm]

\tikzstyle{point_my} = [draw=none, minimum width=0cm, text width=0cm, fill=none, 
  text centered,  minimum height=0cm]    
\tikzstyle{line_my} = [draw, -latex]    
\tikzstyle{box_my}=[draw, minimum size=2em, text width=4.5em, text centered]
\tikzstyle{bigbox_my}=[draw, inner sep=15pt]
\tikzstyle{arrow_my} = [thick,->,>=stealth]
\tikzstyle{noarrow_my} = [thick,-,=>stealth]

\makeatletter
\newcommand\fs@norules{\def\@fs@cfont{\bfseries}\let\@fs@capt\floatc@ruled
  \def\@fs@pre{}%
  \def\@fs@post{}%
  \def\@fs@mid{\kern3pt}%
  \let\@fs@iftopcapt\iftrue}
\makeatother
\restylefloat{algorithm}

\newcommand{\siz}{0.94}  
\newcommand{\sizf}{0.7} 
\newcommand{\sizfs}{0.7}   

\usepackage[acronym,nonumberlist]{glossaries} 
\usepackage{hyperref}
\newacronym[\glsshortpluralkey=PDFs,\glslongpluralkey=probability density functions]{pdf}{PDF}{probability density function}
\newacronym[\glsshortpluralkey=CDFs,\glslongpluralkey=cumulative density functions]{cdf}{CDF}{cumulative density function}
\newacronym[\glsshortpluralkey=CCDFs,\glslongpluralkey=complementary cumulative density functions]{ccdf}{CDF}{complementary cumulative density function}
\newacronym[\glsshortpluralkey=PMFs,\glslongpluralkey=probability mass functions]{pmf}{PMF}{probability mass function}
\newacronym[]{lhs}{l.h.s.}{left-hand side}
\newacronym[]{rhs}{r.h.s.}{right-hand side} 

\newacronym[]{bicm}{BICM}{bit-interleaved coded modulation}
\newacronym[]{bicmid}{BICM-ID}{BICM with iterative demapping}
\newacronym[]{cm}{CM}{coded modulation}
\newacronym[]{tcm}{TCM}{trellis-coded modulation}
\newacronym[]{mlc}{MLC}{multi-level coding}
\newacronym[]{pam}{PAM}{pulse amplitude modulation}
\newacronym[]{bpsk}{BPSK}{binary phase shift keying}
\newacronym[]{qam}{QAM}{quadrature amplitude modulation}
\newacronym[]{16qam}{16-QAM}{16-points quadrature amplitude modulation}
\newacronym[]{psk}{PSK}{phase shift keying}
\newacronym[\glsshortpluralkey=LLRs,\glslongpluralkey=logarithmic likelihood ratios]{llr}{LLR}{logarithmic likelihood ratio}
\newacronym[]{oc}{OC}{operating characteristic}
\newacronym[]{map}{MAP}{maximum a posteriori}
\newacronym[]{ml}{ML}{maximum likelihood}
\newacronym[]{dmp}{DMP}{discretized message passing}
\newacronym[]{mp}{MP}{message passing}
\newacronym[]{ep}{EP}{expectation propagation}
\newacronym[\glsshortpluralkey=MIs,\glslongpluralkey=mutual informations]{mi}{MI}{mutual information}
\newacronym[\glsshortpluralkey=GMIs,\glslongpluralkey=generalized mutual informations]{gmi}{GMI}{generalized mutual information}
\newacronym[]{eesm}{EESM}{exponential effective-SNR-mapping}
\newacronym[]{bicm-gmi}{BICM-GMI}{BICM generalized mutual information}
\newacronym[]{awgn}{AWGN}{additive white Gaussian noise}
\newacronym[]{bsc}{BSC}{binary symetric channel}
\newacronym[]{amc}{AMC}{adaptive modulation and coding}
\newacronym[]{csi}{CSI}{channel state information}
\newacronym[]{cqi}{CQI}{channel quality indicator}
\newacronym[]{kl}{KL}{Kullback-Leibler}
\newacronym[]{cmm}{CMM}{circular moment matching}
\newacronym[]{ga}{GA}{Gaussian approximation}

\newacronym[]{sp}{SP}{set-partitioning}
\newacronym[]{gsm}{GSM}{global system for mobile communications}
\newacronym[]{edge}{EDGE}{enhanced data rates for GSM evolution}
\newacronym[]{3gpp}{3GPP}{3rd generation partnership project}
\newacronym[]{umts}{UMTS}{Universal Mobile Telecommunication System}
\newacronym[]{lte}{LTE}{Long Term Evolution}
\newacronym[]{dvb}{DVB}{digital video broadcasting}
\newacronym[]{fdd}{FDD}{Frequency Division Duplexing}

\newacronym[\glsshortpluralkey=CCs,\glslongpluralkey=convolutional codes]{cc}{CC}{convolutional code}
\newacronym[\glsshortpluralkey=PCCCs,\glslongpluralkey=parallel concatenated convolutional codes]{pccc}{PCCC}{parallel concatenated convolutional code}
\newacronym[\glsshortpluralkey=TCs,\glslongpluralkey=turbo codes]{tc}{TC}{turbo code}
\newacronym{ldpc}{LDPC}{low-density parity-check}
\newacronym[]{ofdm}{OFDM}{orthogonal frequency-division multiplexing}
\newacronym[]{bep}{BEP}{bit-error probability}
\newacronym[]{wep}{WEP}{word-error probability}
\newacronym[]{sep}{SEP}{symbol-error probability}
\newacronym[]{pep}{PEP}{pairwise-error probability}
\newacronym[]{ttcm}{TTCM}{turbo-trellis coded modulation}
\newacronym[]{uep}{UEP}{unequal error protection}
\newacronym[\glsshortpluralkey=CENCs,\glslongpluralkey=convolutional encoders]{cenc}{CENC}{convolutional encoder}
\newacronym[]{mimo}{MIMO}{multiple-input multiple-output}
\newacronym[\glsshortpluralkey=SNRs,\glslongpluralkey=signal-to-noise ratios]{snr}{SNR}{signal-to-noise ratio}
\newacronym[\glsshortpluralkey=SINRs,\glslongpluralkey=the signal-to-interference-plus-noise ratios]{sinr}{SINR}{the signal-to-interference-plus-noise ratio}
\newacronym[]{msb}{MSB}{most-significative bit}
\newacronym[]{bcjr}{BCJR}{Bahl--Cocke--Jelinek--Raviv}
\newacronym[]{cbc}{CBC}{Colavolpe--Barbieri--Caire}
\newacronym[]{skr}{SKR}{Shayovitz--Kreimer--Raphaeli}
\newacronym[\glsshortpluralkey=SEDs,\glslongpluralkey=squared Euclidean distances]{sed}{SED}{squared Euclidean distance}
\newacronym[\glsshortpluralkey=EDs,\glslongpluralkey=Euclidean distances]{ed}{ED}{Euclidean distance}
\newacronym[\glsshortpluralkey=MEDs,\glslongpluralkey=minimum Euclidean distances]{med}{MED}{minimum Euclidean distance}
\newacronym[]{core}{CoRe}{constellation rearrangement}
\newacronym[]{msd}{MSD}{multistage decoding}
\newacronym[]{pdl}{PDL}{parallel decoding of the individual levels}
\newacronym[\glsshortpluralkey=GCs,\glslongpluralkey=Gray codes]{gc}{GC}{Gray code}
\newacronym[]{brgc}{BRGC}{binary-reflected Gray code}
\newacronym[]{nbc}{NBC}{natural binary code}
\newacronym[]{fbc}{FBC}{folded-binary code}
\newacronym[]{bsgc}{BSGC}{binary semi-Gray code}
\newacronym[]{msp}{MSP}{modified set-partitioning}
\newacronym[]{ssp}{SSP}{semi set-partitioning}
\newacronym[]{fhd}{FHD}{free Hamming distance}
\newacronym[]{mfhd}{MFHD}{maximum free Hamming distance}
\newacronym[]{ods}{ODS}{optimal distance spectrum}
\newacronym[]{iud}{i.u.d.}{independent and uniformly distributed}
\newacronym[]{ud}{u.d.}{uniformly distributed}
\newacronym[]{iid}{i.i.d.}{independent, identically distributed}
\newacronym[]{ami}{AMI}{accumulated mutual information}
\newacronym[]{bico}{BICO}{binary-input continuous-output}
\newacronym[]{gh}{GH}{Gauss--Hermite}
\newacronym[]{gum}{GUM}{Gaussian--uniform mixture}

\newacronym[\glsshortpluralkey=BSs,\glslongpluralkey=base-stations]{bs}{BS}{base-station}
\newacronym[\glsshortpluralkey=MSs,\glslongpluralkey=mobile-stations]{ms}{MS}{mobile-stations}

\newacronym[]{phy}{PHY}{physical layer} 
\newacronym[]{rlc}{RLC}{Radio-Link control} 
\newacronym[]{ran}{RAN}{Radio Access Network} 
\newacronym[]{llc}{LLC}{logical link control} 
\newacronym[]{tcp}{TCP}{transmission control protocol} 
\newacronym[]{mac}{MAC}{media access control} 
\newacronym[]{fft}{FFT}{fast Fourier transform} 
\newacronym[]{ft}{FT}{Fourrier transform}
\newacronym[]{cf}{CF}{characteristic function} 
\newacronym[]{mgf}{MGF}{moment generating function} 
\newacronym[]{ee}{EE}{energy efficiency} 
\newacronym[]{eb}{EB}{energy per bit}
\newacronym[]{kkt}{KKT}{Karush--Kuhn--Tucker} 
\newacronym[]{mcs}{MCS}{modulation/coding scheme} 
\newacronym[]{fec}{FEC}{forward error correction}
\newacronym[]{arq}{ARQ}{automatic repeat request}
\newacronym[]{harq}{HARQ}{hybrid ARQ}
\newacronym[]{tarq}{TARQ}{truncated HARQ}
\newacronym[]{ir}{IR}{incremental redundancy}
\newacronym[]{rpr}{RR}{repetition redundancy}
\newacronym[]{rrharq}{RR-HARQ}{repetition redundancy HARQ}
\newacronym[]{irharq}{IR-HARQ}{incremental redundancy HARQ}
\newacronym[]{ack}{ACK}{positive acknowledgment}
\newacronym[]{nack}{NACK}{negative acknowledgment}
\newacronym[]{hol}{HoL}{head of the line}
\newacronym[]{crc}{CRC}{cyclic redundancy check}
\newacronym[]{dp}{DP}{dynamic programming}
\newacronym[]{gp}{GP}{geometric programming}
\newacronym[]{per}{PER}{packet error rate}
\newacronym[]{ber}{BER}{bit error rate}
\newacronym[]{op}{OP}{outage probability}
\newacronym[]{spa}{SPA}{saddle-point approximation}
\newacronym[]{mrc}{MRC}{maximum ratio combining}
\newacronym[]{mdp}{MDP}{Markov decision process}
\newacronym[]{lp}{LP}{linear programming}
\newacronym[]{pomdp}{POMDP}{partially observable Markov decision process}
\newacronym[]{psimdp}{PSI-MDP}{partial state information Markov decision process}
\newacronym[]{scpp}{SCPP}{stochastic shortest path problem}

\newacronym[]{forw}{frwd}{forward}
\newacronym[]{feed}{fdbk}{feedback}

\newacronym[]{mm}{MM-HARQ}{multi-message HARQ}
\newacronym[]{xp}{XP-HARQ}{cross-packet HARQ}
\newacronym[]{ts}{TS}{time-sharing}
\newacronym[]{sc}{SC}{superposition coding}
\newacronym[]{sbrq}{SBRQ}{systematic backward retransmission}
\newacronym[]{brq}{BRQ}{backward retransmission}
\newacronym[]{lharq}{L-HARQ}{layer-coded HARQ}
\newacronym[]{anlharq}{AoN-HARQ}{all-or-none L-HARQ}
\newacronym[]{vlharq}{VL-HARQ}{variable-length HARQ}

\newacronym[]{pp}{PPP}{point process}
\newacronym[]{ppp}{PPP}{Poisson point process}
\newacronym[]{pgfl}{PGFL}{Poisson point process}

\newacronym[]{fide}{FIDE}{F\'ed\'eration Internationale des \'Echecs}
\newacronym[]{fifa}{FIFA}{F\'ed\'eration Internationale de Football Association}
\newacronym[]{epl}{EPL}{English Premier Ligue}
\newacronym[]{nhl}{NHL}{National Hockey Ligue}
\newacronym[]{sg}{SG}{stochastic gradient}
\newacronym[]{lms}{LMS}{least mean squares}
\newacronym[]{rls}{RLS}{recursive least squares}
\newacronym[]{vss}{VSS}{variable step-size}
\newacronym[]{hfa}{HFA}{home-field advantage}
\newacronym[]{mov}{MOV}{margin of victory}
\newacronym[]{ac}{AC}{adjacent categories}

\newacronym[]{tpb}{TPB}{tensor-product-basis}

\usepackage{tikz,pgfplots}
\pgfplotsset{width=3.4in,height=2.3in,compat=1.12} 
\pgfplotsset{width=3.4in,height=2.3in} 
\usetikzlibrary{positioning,shadows,shapes,fit,arrows,backgrounds}  

\usepackage{balance}

\newtheorem{definition}{Definition}

\newtheorem{example}{Example}

\usetikzlibrary{positioning,arrows.meta,bending}
\begin{document}

\title{Parametric Phase Tracking \\ via Expectation Propagation}
\author{Leszek Szczecinski, Hsan Bouazizi, and Ahikam Aharony
\thanks{%
L. Szczecinski is with 
INRS-EMT, University of Quebec, Montreal, Canada. e-mail: leszek@emt.inrs.ca}
\thanks{%
H. Bouazizi was with
INRS-EMT, University of Quebec, Montreal, Canada. e-mail: bouazizi@emt.inrs.ca}
\thanks{%
A. Aharony is with DRW, Chicago, US. e-mail: Ahikam@gmail.com}
}%


\maketitle
\thispagestyle{empty}

\begin{abstract}
In this work we propose simple algorithms for signal detection in a single-carrier transmission corrupted by a strong phase noise. The proposed phase tracking algorithms are formulated within the framework of a parametric message passing (MP) which reduces the complexity of the Bayesian inference by using distributions from a predefined family; here, of Tikhonov distributions. This stays in line with previous works mainly inspired by the well-known \acrfull{cbc} algorithm which gained popularity due to its simplicity and possibility for decoder-aided operation. In our work we mainly focus on practically relevant case of one-shot phase tracking that does not require decoder' feedback. Applying the principles of the \acrfull{ep}, we notably improve the performance of the phase tracking before the decoder's feedback can be even considered. The  \acrshort{ep} algorithms can be also integrated in the decoding loop in the spirit of joint decoding and phase tracking.
	
\end{abstract} 

\section{Introduction}

In this work we propose and analyze simple algorithms for signal detection in the presence of  the phase noise. 

We consider the single-carrier transmission which is often used for high speed communication in frequency non-selective channels \eg in satellite channels \cite{Colavolpe05}, wireless backhaul \cite{ShahMohammadian18}, or optical communications \cite{Millar16,Alfredsson19}. The limiting factor in achieving a high spectral efficiency is then not only due to the presence of the \gls{awgn} but also due to the phase-noise which is caused by the instability of the phase reference (such as a laser or an RF oscillator). 

Although sometimes the concept of ``strong'' phase noise is used, it merely depends on the constellation size: with sufficiently large modulation, the phase noise becomes always a limiting factor for reliable communications. It is especially true for the wireless backhaul and optical channel where the modulation size is aggressively increased \cite{Millar16,ShahMohammadian18,Alfredsson19}. The wireless backhaul, for example, calls for the use of constellation which may contain thousands of points \cite{ShahMohammadian18}. 

Since the phase noise is a random process with memory, the term ``phase tracking'' is often used and a common strategy is to use pilots symbols which, providing reliable reference, facilitate the tracking of the phase for payload symbols.  Increasing the density of the pilots, usually improves the performance  at the cost of decreased spectral efficiency. Therefore, the challenge of the phase tracking is not to eliminate the pilots altogether but rather to attain desirable performance (\eg as measured by the errors rates) with the limited number of pilots. 

The problem of phase tracking can be conveniently formulated using the graph relating all the involved variables \cite{Dauwels04,Colavolpe05}, which also leads to optimal solution defined by the \gls{mp} defined over the graph. It allows us to track the distribution of the phase at each symbol but, to represent accurately the distributions, a large number of samples may be required \cite{Colavolpe05,Shayovitz16} even if  simplifications may be sought by truncating the support of the distributions \cite{ShahMohammadian18}.

Therefore, many works looked into the possibility of \emph{parametric} phase tracking, where the distributions of the phase is assumed to belong to a predefined family;  Tikhonov distributions \cite{Colavolpe05,Shayovitz16} or Gaussians \cite{Kreimer18,Alfredsson19} are the most popular choices. Then, instead of estimating the samples of the distributions, only a few parameters need to be tracked. 

The distributions in the \gls{mp} are naturally mixtures (result of  averaging the contribution of the unknown payload symbols over the entire constellation), with potentially large number of elements. The mixture reduction is thus at the heart of any parametric phase tracking, and consists in replacing the mixture with a member from the adopted family of distributions. In this work we opt for the Tikhonov family of distributions which appear naturally in the context of the phase tracking. This allows us to focus on the principles of the \gls{mp}; we discuss this issue in more details in \secref{Sec:Parametric.MP}.  

In most works the mixture reduction consists in replacing the mixture with one  distribution. The notable difference is \cite{Shayovitz16} which proposes to replace the mixture with another mixture but which contains a small number of elements. The improvements are thus obtained by making the parametrization  more involved both in terms of the number of required parameters and of their estimation.

This is a  general feature of the parametric/approximate \gls{mp} algorithms: the challenge resides in finding a right balance between their performance and the implementation complexity. 

With that regard, the \gls{cbc} algorithm \cite{Colavolpe05} makes uttermost simplification: the mixture reduction is performed \emph{before} the \gls{mp} recursion is invoked. This yields a very simple \gls{mp} algorithm which, due to its simplicity and ensuing popularity, should be considered ``canonical''. However, the cost is paid with relatively poor performance which must be improved  by leveraging the presence of the decoder as only in this way the \gls{cbc} algorithm can exploit the information about the modulation constellation. The resulting joint phase tracking and decoding  relies thus on the iterative exchange of information between the phase tracking and the decoder. This approach has been often reused, \eg \cite{Pecorino15,Alfredsson19}.

It is not without the pitfalls, however, and it was shown in \cite{Kreimer18} that, for large coding rates the \gls{cbc} algorithm introduces an error floor in high \gls{snr}. In our work we will also implement the \gls{cbc} algorithm, observe the error floor effect, and remove it by a scaling down of the unreliable \glspl{llr} delivered by the decoder. 

Even with this improvement, the \gls{cbc} algorithm is failing to approach the limits defined by the \gls{dmp} algorithm  with moderate number of decoding iterations. In fact,  the original work, \cite{Colavolpe05}, considered hundreds  of decoding iterations, which is not the common solution in the current industrial practice where ten(s) of iterations are rather preferred.

To remedy the poor performance of the \gls{cbc}, the mixture reduction may be done \emph{within} the recursive equations of the \gls{mp}. This is the idea of the algorithms proposed in \cite{Shayovitz16,Kreimer18}  which, even in the absence of the decoder's feedback are able to exploit the form of the constellation; this improves the performance at the cost of more complex \gls{mp}. 

What the \gls{cbc} algorithm and many works on parametric \gls{mp}, \eg \cite{Shayovitz16,Kreimer18}, have in common is that they rely heavily on the decoder to improve the estimation of the phase via joint decoding - phase tracking. And while it is clear that such a joint operation improves the performance, much less attention was  paid to the performance of ``one-shot'' phase tracking (meaning that the latter is decoupled from the decoder's outcome). This is a relevant issue mainly because it is compatible with a current industrial practice\footnote{To our best knowledge, the phase tracking algorithms used in industrial products do not use decoder's feedback.} but also because efficient one-shot algorithm should yield larger improvements when operating jointly with the decoder.

The above considerations explain the main motivation behind our work:  we want to exploit the form of the constellation \emph{without} relying on the decoder's feedback.

We first note that the \gls{cbc} algorithm as well as those proposed in \cite{Shayovitz16,Kreimer18}, relying on approximations (due to the mixture reduction),  are  suboptimal by nature. Our main idea is to improve them  iteratively using the \gls{ep} \cite{Minka01} which is a general framework for iterative refinement of the approximations in parametric \gls{mp}s. 

We show that the \gls{ep} improves significantly the performance of one-shot receivers closing notably the performance gap to the \gls{dmp}. This advantage also materializes in the iterative versions of the proposed algorithms.

The rest of the paper is organized as follows. In~\secref{sec:model}, we introduce the adopted system model, in \secref{sec:phase.track} we outline the fundamentals of the Bayesian phase tracking while its parametric formulation is explained in \secref{Sec:Parametric.MP}.

The main contribution lies in \secref{Sec:It.EP} which explains how the \gls{ep} framework may be used to derive a new phase tracking algorithms which, as shown in numerical examples, improve the performance without any help from the decoder. We also show that, when combined with the decoder, the new algorithms approach closely the  \gls{dmp} limits. The conclusions are drawn in \secref{sec:Conclusions} while the appendices show the details of the operations on the circular distributions and provide new approximations required in the  latter.

\section{System Model}\label{sec:model}

We consider transmission over a channel corrupted by the additive noise and phased noise 
\begin{align}\label{y.kn}
y_{n} =  x_{n}\e^{\jj \theta_{n}} + v_{n},\quad n=0,\ld, N
\end{align}
where $x_n\in\mcA$ is the transmitted complex symbol drawn from the $M$-ary constellation $\mcA$, \ie $|\mcA|=M$, $y_{n}$ are samples of the received signal, $v_{n}$ is the additive noise, and $\theta_n$ is the phase noise. 

We assume the constellation $\mcA$ is zero mean and energy normalized so, modelling $x_n$ as random variables obtained by uniform sampling of  $\mcA$, we have $\Ex[x_n]=0$ and \mbox{$\Ex[x^2_n]=1$}. We model $v_{n}$ as complex \gls{iid} Gaussian variables with zero mean and variance $\No$ (\gls{awgn} model); $\theta_{n}$ is modelled as a Wiener process
\begin{align}\label{theta.Wiener}
\theta_{n}=\theta_{n-1}+w_{n}, \quad n=1,\ld,N,
\end{align}
where $w_{n}$ are \gls{iid}  zero-mean, real Gaussian variables with variance $\sigma_w^2$; the initial value $\theta_{0}$  is modelled as uniformly distributed over the interval $(-\pi, \pi]$.

This model is popular in wireless and optical communications, \eg \cite{Colavolpe05,Alfredsson19}. The variance of the additive noise $v_n$ is determined by the thermal/optical noise at the receiver and the attenuation on the propagation path; the \gls{snr} is defined as $\SNRrv=\frac{1}{\No}$. The variance of the phase noise $w_{n}$ reflects the (in)stability of the oscillator used in the demodulation process. We suppose that both, $\sigma_w^2$ and $\SNRrv$, are known at the receiver. 

The transmitted symbols $x_n$ are obtained via \gls{bicm} \cite[Chap.~1.4]{Szczecinski_Alvarado_book} in two steps: 1) the information bits $\set{b_n}_{n=1}^{\Nb}$ are encoded using the binary encoder of rate $r$ into the coded bits $\set{c_n}_{n=1}^{\Nc}$, where $\Nb=r\Nc$ (in the numerical examples we use the \gls{ldpc} codes, and 2) the coded bits $\set{c_n}$ are regrouped into length-$m$ labels $\tilde{\bc}_n=[c_{n,1},\ld, c_{n,m}], n=1,\ld, \Ns$, where $\Ns m=\Nc$; they are next mapped onto the symbols from the constellation $\mcA$
\begin{align}\label{s.n.c}
s_{n}&=\MODwo\big[\tilde{\bc}_n\big], \quad n=1,\ld, \Ns,
\end{align} 
where $\MODwo[\cd]: \set{0,1}^m\mapsto \mcA$ defines the mapping; in this work, $\mcA$ is $M$-ary \gls{qam} and $\MODwo[\cd]$ is the Gray mapping \cite[Sec.~2.5.2]{Szczecinski_Alvarado_book}.
Finally, reference symbols (pilots) are interleaved with the payload $\set{s_n}_{n=1}^\Ns$ so the transmitted sequence can be presented as
\begin{align}
\label{payload.1}
\set{x_n}_{n=1}^{N}
&=\set{ \underset{\uparrow}{x_0},x_1,\ld, x_{L-1},\underset{\uparrow}{x_L},x_{L+1},\ld, \underset{\uparrow}{x_N} },
\end{align}
where we indicate with arrows the pilot symbols $x_{n}, n\in\Natural_{\tr{pilots}}$, while $x_n, n\in\Natural_{\tr{payload}}$ are payload symbols with the sets of indices to the payload symbols and to the pilots defined as
\begin{align}
\Natural_{\tr{pilots}}&=\set{0,L,2L,\ld, FL}\\
\Natural_{\tr{payload}}&=\set{1,\ld, N} \setminus \Natural_{\tr{pilots}}.
\end{align}
We will use the values of $L$ which satisfy $\Ns=F (L-1)$ for integer $F$.  We also use the mapping $n'=n'(n)$ which allows us to index the symbols in $\set{x_n}$ by skipping the pilots, that is, $x_{n'}=s_n, n=1,\ld,\Ns$.

Similarly, the \gls{bicm} decoding is carried out  in two steps \cite[Chap.~1.4]{Szczecinski_Alvarado_book}
\begin{enumerate}
\item \emph{Demodulation}: consists in finding the marginal conditional probability of the coded bits
\begin{align}\label{p.nk.c}
&\PR{c_{n,k}=c|\by}\\
\label{demodulation}
&\qquad ~ \quad 
\propto \sum_{a\in\mcA_{k,c}}\PR{s_{n}=a|\by},
\end{align}
where $\by=\set{y_n}_{n=1}^N$ gathers all the channel outcomes and 
\begin{align}
\mcA_{k,c}=\set{ \MOD{\bc}, \bc=[c_1,\ld, c_{k-1},c,c_{k+1},\ld, c_m]}
\end{align}
is a sub-constellation comprising only the symbols labeled by the bit with value $c\in\set{0,1}$ at the position $k=1,\ld,m$;  $\propto$ will be used in the text to indicate that the distributions are defined up to a multiplicative factor which is independent of the distribution argument (here $c$ or $a$). 


Since $s_n=x_{n'}$, to obtain \eqref{p.nk.c} we have to calculate the conditional distribution of the symbols $x_n$
\begin{align}\label{P.n.a.x}
P_n(a)\triangleq\PR{x_{n}=a|\by}, \quad a\in\mcA;
\end{align} 
this operation involves marginalization over the phase $\theta_{n}$ and the transmitted symbols $x_l, l\neq n$, and is a ``phase tracking'' since, as a byproduct of \eqref{P.n.a.x}, we will obtain the distribution of the phase $\pdf(\theta_n|\set{y_l}_{l=1}^N)$.

\item \emph{Soft-input decoding}: using \eqref{P.n.a.x} and \eqref{p.nk.c} we calculate the \glspl{llr} for the coded bits $c_{n,k}$, \ie 
\begin{align}\nonumber
\lambda_{n,k}&=\log\frac{\PR{c_{n,k}=1|\by}}{\PR{c_{n,k}=0|\by}}\\
\label{lambda.n.l}
&\approx
\max_{a\in\mcA_{k,1}}\hat{P}_{n'}(a) - \max_{a\in\mcA_{k,0}}\hat{P}_{n'}(a),
\end{align}
where we applied the max-log simplification using the log-probability, $\hat{P}_n(a)=\log P_n(a)$; the logarithmic domain eases implementation and will appear in all parametric derivations; we use again the mapping $n'=n'(n)$. 
\end{enumerate}

The \gls{llr}s, $\lambda_{n,k}$,  are fed to the binary decoder which operates in abstraction of how they were  calculated. This very separation of the operation of the decoder and the demodulator is the distinctive feature of the \gls{bicm} which made it, de facto, a standard approach to design spectrally efficient transceivers. The obvious advantage is that both, the demodulator (comprising the phase-tracking) and the decoder may be designed and implemented independently of each other. Such an operation is characteristic of ``one-shot'' demodulators which are the most common solutions in the \gls{bicm} transceivers. 

On the other hand, the inherent simplicity of one-shot demodulators is a source of performance limitation which may be overcome by forcing the two-way exchange between the demodulator and the decoder. We also consider this option and we will assume that the decoder provides the demodulator with the prior \glspl{llr}, $\lambda^\tr{a}_{n,k}$, for the bits $c_{n,k}$ from which the prior symbol probabilities $P^\tr{a}_n(a)=\PR{x_n=a}$ are calculated as
\begin{align}\label{Pa.lambda}
\hat{P}^\tr{a}_n(a)=\log P^\tr{a}_n(a)  \propto \sum_{k=1}^m \lambda^\tr{a}_{n,k} \tr{bit}_{k}(a)
\end{align}
where $\tr{bit}_{k}(a)$ is the value of the $k$-th bit in the label of the symbol $a$.

If we opt for such an iterative demodulation/phase-tracking, we will nevetheless preserve the \gls{llr} calculation in \eqref{lambda.n.l} which does not take advantage of the \glspl{llr},  $\lambda^\tr{a}_{n,k}$.\footnote{It is possible, however. This approach is known to provides little gain in the case of the Gray-mapped constellations we use here \cite{Schreckenbach03} but yields gains with appropriately designed mapping \cite{Szczecinski05e}.}

\section{Phase tracking}\label{sec:phase.track}
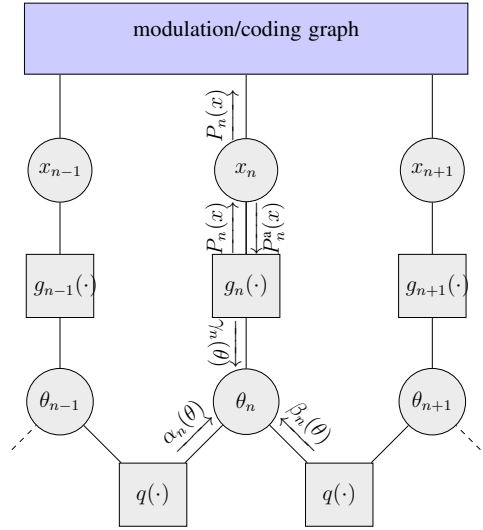
\begin{figure}[bt]
\begin{center}
\scalebox{\sizf}{
\pgfdeclarelayer{background}
\pgfdeclarelayer{foreground}
\pgfsetlayers{background,main,foreground}


\begin{tikzpicture}[trim left=0cm,node distance=3cm]

\node[circle_my_graph] 	(phase11) 		{$\theta_{n-1}$};
\node[square_my_graph]	(Chan11) 	[above of=phase11,node distance=2.2cm]	{$g_{n-1}(\cd)$};
\node[circle_my_graph] 	(symbol11) 	[above of=Chan11,node distance=2.2cm]	{$x_{n-1}$};

\node[square_my_graph]	(Pw1) 	[below right of=phase11,node distance=2.5cm]	{$q(\cd)$};


\node[point_my]	 		(point11) 	[below left of=phase11,node distance=1.5cm]	{};
\draw[-]   (Pw1)	  	--   (phase11);  
\draw[-]   (phase11)  -- 	(Chan11);  
\draw[-]   (Chan11) -- (symbol11);  
\draw[-,dashed]   (point11)	  -- 	(phase11);   


\node[circle_my_graph] 	(phase1) 	[above right of=Pw1,node distance=2.5cm]	{$\theta_{n}$};
\node[square_my_graph]	(Chan1) 	[above of=phase1,node distance=2.2cm]	{$g_{n}(\cd)$};
\node[circle_my_graph] 	(symbol1) 	[above of=Chan1,node distance=2.2cm]	{$x_{n}$};

\draw[-]   (Pw1)	  --  node[midway,above,sloped]{$\underrightarrow{\alpha_{n}(\theta)}$}	(phase1);   
\draw[-]   (Chan1) -- node[midway,below,sloped]{$\overrightarrow{\gamma_{n}(\theta)}$} (phase1);   
\draw[-]   (Chan1) -- node[midway,above,sloped]{$\underrightarrow{P_{n}(x)}$} (symbol1);  
\draw[-]   (Chan1) -- node[midway,below,sloped]{$\overleftarrow{P^{\tr{a}}_{n}(x)}$} (symbol1);  

\node[square_my_graph]	(Pw13) 	[below right of=phase1,node distance=2.5cm]	{$q(\cd)$};
\node[circle_my_graph] 	(phase13) 	[above right of=Pw13,node distance=2.5cm]	{$\theta_{n+1}$};
\node[square_my_graph]	(Chan13) 	[above of=phase13,node distance=2.2cm]	{$g_{n+1}(\cd)$};
\node[circle_my_graph] 	(symbol13) 	[above of=Chan13,node distance=2.2cm]	{$x_{n+1}$};
\node[point_my]	 		(point13) 	[below right of=phase13,node distance=1.5cm]	{};
\draw[-]   (Pw13)	  -- node[midway,above,sloped]{$\underleftarrow{\beta_{n}(\theta)}$}	(phase1);   
\draw[-]   (Pw13)	  -- 	(phase13);   
\draw[-]   (phase13)  -- 	(Chan13);   
\draw[-]   (Chan13) -- (symbol13);  
\draw[-,dashed]   (point13)	  -- 	(phase13);   

\begin{pgfonlayer}{foreground}
\node[point_my]	 (label11H)	[above of=symbol11,node distance=2.5cm]	{};
\node[point_my]	 (label1H)	[above of=symbol1,node distance=2.5cm]	{};
\node[point_my]	 (label13H)	[above of=symbol13,node distance=2.5cm]	{};

\draw[-]    (symbol1) -- node[near start,above,sloped]{$\underrightarrow{P_{n}(x)}$} (label1H);   
\draw[-]    (symbol11) -- (label11H);   
\draw[-]    (symbol13) --  (label13H);  
\node[bigbox_my,fill=blue!20, fit=(label11H)(label13H)]   (ENCODER) {modulation/coding graph};
\end{pgfonlayer}
 


\end{tikzpicture}}
\end{center}
\caption{Graph defining the relationship between the variables in the phase tracking problem used previously \eg in \cite[Fig.~2]{Colavolpe05}, \cite[Fig.~2]{Shayovitz16}; the messages exchanged between the nodes are shown together with the arrows indicating the direction of the exchange. When variable is connected to two functions, the message passes transparently through the variable node as in the case of probability $P_n(a)=\PR{x_n=a|\by}, a\in\mcA$ defining the distribution of the symbol $x_n$ via phase tracking, and its priori distribution $P_n^{\tr{a}}(a), a\in\mcA$  (obtained from the decoder).}
\label{fig:graph.tracking}
\end{figure}

Our goal now is to find $P_n(a), a\in\mcA$ exploiting the relationships between the involved random variables.  

As in \cite{Colavolpe05,Kreimer18}, we show in \figref{fig:graph.tracking} a graph which captures these relationships: the squares represent the functions taking as arguments the variables which, in turn, are represented by the directly connected circles. In particular
\begin{align}
\label{f.n.x}
g_n(\theta_n,x_n)  & \propto f(y_n|x_n,\theta_n) \\
&\propto \exp\left(-\SNRrv\Big|y_n-x_n\e^{\jmath\theta_n}\Big|^2\right)
\end{align}
corresponds to the relationship defined in \eqref{y.kn}, where we use $\pdf(\cd|\cd)$ to denote the conditional \gls{pdf}; the model \eqref{theta.Wiener} yields 
\begin{align}\label{p.w}
q(\theta_n,\theta_{n+1}) &\propto \pdf(\theta_n|\theta_{n+1})\propto\pdf( \theta_{n+1} | \theta_{n} )\\
&=\omega(\theta_n-\theta_{n+1}),
\end{align}
with
\begin{align}\label{Normal.pdf}
\omega(\theta)=\frac{1}{\sqrt{2\pi \sigma^2_w}}\sum_{k=-\infty}^{\infty}\exp\left(-\frac{(\theta-k2\pi)^2}{2\sigma^2_w}\right)
\end{align}
being a zero-mean, wrapped Gaussian distribution defined over any interval of length $2\pi$.

In \figref{fig:graph.tracking} we also show a shaded rectangle labeled as ``modulation/coding graph''; it contains a graph which describes the relationship $\set{x_n}  \rightarrow \set{s_n} \rightarrow \set{c_n} \rightarrow \set{b_n}$, and its knowledge is used for the demodulation (to implement \eqref{lambda.n.l}) and for the decoding.

If the phase tracking has to be carried out and no decoding was not yet executed,  the payload symbols $x_n$ are assumed to be \gls{iid}, with non-informative a priori distribution over the transmitted symbols, \ie $P^{\tr{a}}_n(a)\propto 1$. In a case we would like  to perform the joint decoding/demodulation/phase tracking, as done, \eg in \cite{Colavolpe05,Kreimer18,Alfredsson19}, we use $P^{\tr{a}}_n(a)$ obtained from the decoding results. 

Thus, from the point of view of the phase tracking, the relationships induced by the modulation/coding graph are summarized by $P^{\tr{a}}_n(a)$. As a direct consequence, the graph we deal with in \figref{fig:graph.tracking} is a tree, \ie contain no loops. Therefore, the efficient marginalization can be done exactly by the \gls{mp} algorithm as follows \cite{Colavolpe05}, \cite{Shayovitz16}, \cite{Kreimer18}
\begin{align}\label{P.n.x}
P_n(a) \propto\int_{-\pi}^{\pi}  \alpha_n(\theta)  \beta_n(\theta) g_n(\theta,a ) \dd \theta,\quad a\in\mcA,
\end{align}
where
\begin{align}\label{alpha}
\alpha_n(\theta)&\propto \big(\alpha_{n-1}(\theta)  \gamma_{n-1}(\theta)\big) \convop 
\omega(\theta),\\
\label{beta}
\beta_n(\theta)&\propto \big(\beta_{n+1}(\theta) \gamma_{n+1}(\theta) \big) \convop 
\omega(\theta),\\
\label{gamma}
\gamma_n(\theta)&\propto\sum_{a\in\mcA} P^{\tr{a}}_n(a) g_{n}(\theta,a ),
\end{align}
are (proportional to) the marginal distributions of the phase $\theta_n$ conditioned, respectively, on $y_0,\ld,y_{n-1}$, on $y_{n+1},\ld,y_{N}$, and  on $y_n$.\footnote{That is,
\begin{align}
\alpha_n(\theta)&\propto\pdf(\theta_n| y_0,\ld,y_{n-1})\\
\beta_n(\theta)&\propto\pdf(\theta_n| y_{n+1},\ld,y_{N})\\
g_n(\theta,a)&\propto\pdf(\theta_n| y_{n},a)
\end{align}
and thus the integrand in \eqref{P.n.x} is the posterior distribution of the phase
\begin{align}
\pdf(\theta_n | \by, a)&\propto
 \alpha_n(\theta)  \beta_n(\theta) g_n(\theta,a ) 
\end{align}
while $P_n(a)$ is given by \eqref{P.n.a.x}.

} All the distributions of the phase are periodic and so is the convolution in \eqref{alpha} and \eqref{beta}. 

\section{Parametric message passing}\label{Sec:Parametric.MP}
The operations \eqref{P.n.x}-\eqref{gamma} which allow us to calculate $P_n(a)$ may be implemented on the discretized versions of the distributions $\alpha_n(\theta)$, $\beta_n(\theta)$, and $\gamma_n(\theta)$ defined over the samples of $\theta\in(-\pi, \pi]$. In such a \acrfull{dmp}, the multiplications and additions are straightforward to implemented and the convolution is efficiently implemented via \gls{fft}. 

However, to ensure sufficient precision of the discretization, a large number of samples, $N_{\theta}$, may be needed.\footnote{\label{foot:N8} The literature often used $N_\theta=8M$ \eg \cite{Peleg00,Colavolpe05,Shayovitz16,Kreimer18} but finding the smallest $N_{\theta}$ which ensures the sufficient accuracy of the \gls{mp}, \eqref{alpha}-\eqref{gamma}, is difficult. 

The general rule is that all the distributions involved in the \gls{mp} should be represented without ``significant'' loss of accuracy so the sampling step should decrease when the spectral support of the distributions increases. 

Then, by approximating the distributions by Gaussians, we know that their spectral support is proportional to their precision (inverse of the variance). Thus, observing that the precision of $g_n(\theta,a)$ increases with the \gls{snr}, see also \eqref{f.n.x.Tikh}-\eqref{z.fn}, and the precision of $\omega(\theta)$ decreases with $\sigma_w^2$, we may conclude that the number of samples will grow when we increase the \gls{snr} or when we decrease the phase-noise level. These operational conditions are also required to transmit reliably high-order constellations. So, indeed, we should expect that the $N_\theta$ will grow with $M$ but we are not aware of any formal study backing up the heuristics $N_\theta=8M$ (that turned out unnecessarily conservative in our case, see Footnote~\ref{foot:DMP32} in \secref{example.start}.).
}  Since this implies a large computational complexity as well as a storage requirements, many previous works have gone in the direction of parametric representation of the involved periodic distributions. Then, instead of tracking the samples in \gls{dmp}, we only need to track (a few) parameters which represent the distributions. 

We thus should find a family of the distributions, $\mcF$, on which the operations defined in \eqref{P.n.x}-\eqref{gamma} are easily dealt with. In particular, we have to handle multiplication of the periodic functions as well as, the convolution in \eqref{alpha} and \eqref{beta}; the integration in \eqref{P.n.x}, when done on the parametric forms of distributions, is usually straightforward.

In this work we consider a family of  Tikhonov distributions which were already used, \eg in \cite{Colavolpe05,Barbieri07,Shayovitz16}. Here, to simplify the notation we use its unnormalized version 
\begin{align}\label{Tikh.definition}
\mcT(\theta; z)= \e^{\Re[z\e^{-\jj \theta}]},
\end{align}
where $\Re[\cd]$ denotes the real part and $z$ is a complex parameter whose phase is the circular mean of the underlying variable and the value of $|z|$ is, approximatively, a measure of precision \cite{Kurz16}.\footnote{The normalized distribution is obtained as
\begin{align}
\mcT^\text{norm}(\theta; z)=\frac{\mcT(\theta; z)}{\int_0^{2\pi}\mcT(\theta; z)\dd \theta}=\frac{\e^{\Re[z\e^{-\jj \theta}]}}{2\pi I_0(|z|)},
\end{align}
where $I_0(\cd)$ is the zero-th order modified Bessel function. 
}

The multiplications are trivial over $\mcF$ 
\begin{align}\label{Tikh.multiplication}
\mcT(\theta; z_1)\mcT(\theta; z_2)
&=
\mcT(\theta; z_1+z_2),
\end{align}
while the convolution with the Gaussian, appearing in \eqref{alpha} and \eqref{beta}, must be treated via  approximations as done before \cite[Appendix]{Colavolpe05} and also explained using the concept of a \gls{cmm} in \appref{conv.T.G}
\begin{align}\label{Tik.h.Normal.convolution}
\mcT(\theta; z)&\convop\omega(\theta)
\propto  \mcT\Big( \theta; \frac{z}{1+\sigma_w^2|z|} \Big).
\end{align}

We also note that $g_n(\theta,a) \in \mcF$, so \eqref{f.n.x} can be rewritten as
\begin{align}\label{f.n.x.Tikh}
g_n(\theta,a)&
\propto \e^{-\SNRrv |a|^2}\mcT\big(\theta; z_{g,n}(a)\big) \\
\label{z.fn}
z_{g,n}(a)&=2\SNRrv ~y_n a^{*}.
\end{align}

Instead of Tikhonov family we might opt for the wrapped Gaussians \eqref{Normal.pdf} but 
\begin{itemize}
\item The family of circular Gaussian distributions is not closed under multiplications \cite[Sec.~II.B]{Kurz16} so approximations are needed. This aspect is not always emphasized in the literature: in some works,  \eg in \cite{Kreimer18}, the Gaussian distributions are used without explicit wrapping. Ignoring the circularity, it produces an apparent closeness under multiplication but may lead to interpretation errors when ``locking'' on the wrong phase, see example in \cite[Fig.~1]{Kurz16}; 
\item We have to start using approximations already for $g_n(\theta,a)$; it can be done, of course, \eg as in \cite{Kreimer18}, but requires the introduction of additional layer of approximations before the \gls{mp} is derived. This stands in contrast with a natural representation of  $g_n(\theta,a)$ by a Tikhonov distribution in \eqref{f.n.x.Tikh},
and 
\item Comparing to the Tikhonov distributions, Gaussians are less well suited to represent the uniform or almost-uniform distributions of the phase (in Tikhonov family this is obtained by setting $z\approx 0$, while in Gaussian the variance must tend to infinity).
\end{itemize}


We thus see that, irrespectively of the adopted family $\mcF$, approximations are necessary  (to implement multiplication if $\mcF$ is a Gaussian family, or to implement the convolution if $\mcF$ is a Tikhonov family) but this is not the main issue in the parametric \gls{mp}. The principal difficulty stems from the presence of $\gamma_n(\theta)$  which is a mixture of Tikhonov distributions which  we rewrite as
\begin{align}\label{gamma.def.mixture}
\gamma_n(\theta)&\propto\sum_{a\in\mcA}\e^{\eta_{g,n}(a)}\mcT\big(\theta;z_{g,n}(a )\big),\\
\label{eta.f.n}
\eta_{g,n}(a)&=-\SNRrv ~|a|^2+\hat{P}^{\tr{a}}_n(a),
\end{align}
where we again use $\hat{P}^{\tr{a}}_n(a)=\log P^{\tr{a}}_n(a)$ because the logarithmic representation simplifies the implementation.

There are essentially two approaches that have been used in the literature to deal with this issues and we compare them in the following.

\subsection{Out-of recursion projection: CBC algorithm}\label{Sec:Pre-MP}

Taking advantage of the fact that $\gamma_n(\theta)$ is not obtained recursively, its approximation, $\tilde{\gamma}_n(\theta)$ may be obtained \emph{before} the recursive steps of \gls{mp} \eqref{alpha}-\eqref{beta} are executed. This may be seen as the following approximation of the \gls{rhs} of \eqref{alpha}:
\begin{align}
\label{alpha.projection.preMP}
\alpha_n(\theta)&\propto \mcF\left[\big(\alpha_{n-1}(\theta)  \gamma_{n-1}(\theta)\big) \convop  \omega(\theta)\right]\\
\label{alpha.projection.preMP.1}
&\approx \mcF\left[\alpha_{n-1}(\theta)  \gamma_{n-1}(\theta)\right] \convop  \omega(\theta)\\
\label{alpha.projection.preMP.2}
&\approx \big(\alpha_{n-1}(\theta)  \mcF[\gamma_{n-1}(\theta)]\big) \convop  \omega(\theta)\\
\label{alpha.projection.preMP.3}
&= \big(\alpha_{n-1}(\theta)  \tilde\gamma_{n-1}(\theta)\big) \convop  \omega(\theta)
\end{align}
where, 
\begin{align}\label{gamma.projection.first}
\tilde{\gamma}_n(\theta)
&=\mcF[\gamma_n(\theta)]=\mcT\Big(\theta;  z_{\tilde{\gamma},n} \Big)\\
&=\mcF\Big[ \sum_{a\in\mcA}\e^{\eta_{g,n}(a)}\mcT\big(\theta;z_{g,n}(a )\big) \Big],
\end{align}
and, with a slight abuse of notation we use $\mcF[\cd]$ to denote the operator which  projects $\gamma_n(\theta)$ onto the Tikhonov family $\mcF$. To pass from \eqref{alpha.projection.preMP} to \eqref{alpha.projection.preMP.1} we used (approximate) closeness of $\mcF$ with respect to convolution and this approximation holds well. On the other hand, to pass from  \eqref{alpha.projection.preMP.1} to \eqref{alpha.projection.preMP.2} we used the   approximation $\mcF\left[\alpha_{n-1}(\theta)  \gamma_{n-1}(\theta)\right]\approx \alpha_{n-1}(\theta)\mcF\left[\gamma_{n-1}(\theta) \right]$. Since $\gamma_n(\theta)$ is a mixture of Tikhonov distributions, this approximation is very crude but it allows us to precalculate $\tilde{\gamma}_n(\theta)$ before the \gls{mp} recursion starts. This idea underlies the \acrfull{cbc} algorithm introduced in \cite{Colavolpe05}.

Since all the distributions belong then to $\mcF$, \ie $\alpha_n(\theta)=\mcT(\theta;z_{\alpha,n})$ and $\beta_n(\theta)=\mcT(\theta;z_{\beta,n})$, replacing $\gamma_n(\theta)$ with $\tilde{\gamma}_{n}(\theta)$ (as defined in  \eqref{gamma.projection.first})  and applying \eqref{Tikh.multiplication} and \eqref{Tik.h.Normal.convolution}, we obtain the following parametric \gls{mp} :
\begin{align}
\label{z.alpha.init}
z_{\alpha,0}&=0,\\
\label{z.alpha}
z_{\alpha,n}&= \frac{z_{\alpha,n-1} + z_{\tilde{\gamma},n-1}}{1+|z_{\alpha,n-1} + z_{\tilde{\gamma},n-1}|\sigma_w^2}, \quad n=1, \ld,N-1\\
\label{z.beta.init}
z_{\beta,N}&=0,\\
\label{z.beta}
z_{\beta,n}&= \frac{z_{\beta,n+1} + z_{\tilde{\gamma},n+1}}{1+|z_{\beta,n+1} + z_{\tilde{\gamma},n+1}|\sigma_w^2}, \quad n=N-1,\ld,1,
\end{align}
which is exactly equivalent to the \gls{cbc} algorithm defined in \cite[Eq.~(36)-(37)]{Colavolpe05} and  is provided for completeness and as a starting point for the discussion.

Using \eqref{Tikh.multiplication}  and \eqref{f.n.x.Tikh}--\eqref{z.fn} in \eqref{P.n.x}, the ``extrinsic'' symbol probabilities are also calculated as \cite[Eq.~(35)]{Colavolpe05}
\begin{align}\label{P.a.final}
P_n( a )&\propto \int_{0}^{2\pi} \mcT\big(\theta;z_{\Sigma,n}(a)\big)\e^{-\SNRrv|a|^2} \dd \theta,\\
\label{P.a.final.2}
\hat{P}_n( a )&=\log P_n( a )\\
&\propto -\SNRrv |a|^2 + \hat{I}_0\big(|z_{\Sigma,n}(a)|\big),
\end{align}
where
\begin{align}
\label{z.Sigma.n}
z_{\Sigma,n}(a)&=z_{\alpha,n}+z_{\beta,n}+z_{g,n}(a),
\end{align}
and we avoid numerical issues due to the exponential grows of $I_0(\cd)$ by using its log-version
\begin{align}\label{hat.I0}
\hat{I}_0(x) &= \log I_0(x) = x  + \Delta(x),
\end{align}
with $\Delta(x)$ being the log-domain corrective factor.\footnote{The approximation $\Delta(x)\approx -\frac{1}{2}\log(2\pi x)\IND{x>\frac{1}{2\pi}}$ is tight for large $x$ \cite[Eq.~(96)]{Shayovitz16}\cite[Eq.~(18)]{Pecorino15}; we truncate it for small arguments, $x<\frac{1}{2\pi}$, because we know that $\Delta( 0 )=0$. For better accuracy, $\Delta(x)$ may be implemented via lookup table. It is also possible to simply set $\Delta(x)\approx 0$.}

Further,  the log-probability $\hat{P}_n( a )$ should be used in \eqref{lambda.n.l}.  

Due to its simplicity the \gls{cbc} algorithm \eqref{z.alpha.init}-\eqref{z.beta}, should be treated as a ``canonical'' solution to the problem of phase tracking: it is explicit and requires no fine-tuning. This, and its historical precedence,  explain the considerable attention it received up to now.

The remaining issue is the projection in \eqref{gamma.projection.first};  \cite{Colavolpe05} proposed to approximate $\gamma_n(\theta)$ with $\tilde{\gamma}_n(\theta)\in\mcF$ by,  first matching a Gaussian to $\pdf(y_n|\theta_n,a)\propto g_n(\theta,a)$ from which a Tikhonov distribution was derived as \cite[Sec.~IV.B]{Colavolpe05}
\begin{align}
\label{hat.z.gamma}
\tilde{\gamma}_n(\theta)
&=\mcT\Big(\theta;  z_{\tilde{\gamma},n} \Big)\\
\label{gamma.GA.cases}
z_{\tilde{\gamma},n}&=
\begin{cases}
z_{g,n}(x_n)  & \text{if}\quad n\in \Natural_{\tr{pilots}}\\
2\frac{y_n\mfm_n^{*}}{\No+ \mfv_n}, & \text{if}\quad n\in\Natural_{\tr{payload}}
\end{cases}
\end{align}
where 
\begin{align}
\label{m.n.A}
\mfm_n&=\sum_{a\in\mcA}  a P^{\tr{a}}_n(a) ,\quad
\mfv_n=\sum_{a\in\mcA}  |a|^2 P^{\tr{a}}_n(a)   -  |\mfm_n|^2,
\end{align}
are the mean and the variance of $x_n$. 

The approximation is needed only for the payload positions, $n\in\Natural_{\tr{payload}}$. For the pilots $x_n, n\in\Natural_{\tr{pilots}}$, we have $\gamma_n(\theta)=g_{n}(\theta, x_{n} )$ so we can use directly the parameter of the Tikhonov distribution $z_{\tilde{\gamma},n}=z_{g,n}(x_n)$.

This \gls{ga} approach was often reused in the literature, \eg \cite{Barbieri07,Pecorino15,Alfredsson19} as the integral part of the \gls{cbc} algorithm. 

On the other hand, recognizing that the distribution \eqref{gamma.def.mixture} is circular, instead of the \gls{ga}  we may apply the \gls{cmm}. It relies on minimization of the \gls{kl} distance between $\tilde{\gamma}_n(\theta)$ and the mixture $\gamma_n(\theta)$, and changes the way the parameter $z_{\tilde{\gamma},n}$ is calculated, which we express as follows
\begin{align}
\label{gamma.projection.first.cases}
z_{\tilde{\gamma},n}
&=
\begin{cases}
z_{g,n}(x_{n}) &\text{if}\quad n\in\Natural_{\tr{pilots}}\\
\mf{CMM}\Big[ \big\{\eta_{g,n}(a), z_{g,n}(a)\big\}_{a\in\mcA}  \Big]&\text{if}\quad n\in\Natural_{\tr{payload}},
\end{cases}
\end{align}
where the dependence of $z_{\tilde{\gamma},n}$ on  $\eta_{g,n}(a)$ and $z_{g,n}(a)$, indexed by $a\in\mcA$  is defined in closed-form by a function $\mf{CMM}[\cd]$ shown in \appref{Sec:CMM}. 

Although the \gls{cmm} was already used in the context of the phase tracking by \cite{Shayovitz16}, we are not aware of it being applied as a part of the \gls{cbc} algorithm. And while the implementation of the function $\mf{CMM}[\cd]$ is slightly less straightforward\footnote{Approximations of non-linear functions are involved, see \appref{Sec:CMM} and \appref{Sec:Bx}.} than the \gls{ga} shown in \eqref{hat.z.gamma}-\eqref{m.n.A}, it is worthwhile to evaluate the advantage of the former.

An important observation, already made \eg in \cite[Sec.~I]{Shayovitz16} or \cite[Sec.~IV]{Pecorino15}, is that, using a non-informative prior $\hat{P}^{\tr{a}}(a)\propto 0$,  the approximation $\tilde{\gamma}_n(\theta)$ is also non-informative for the payload symbols; that is, \mbox{$z_{\tilde{\gamma},n}=0, n\in\Natural_{\tr{payload}}$}.\footnote{If  the \gls{ga} \eqref{hat.z.gamma}-\eqref{m.n.A}, is used,  this is true, if the constellation $\mcA$ is zero mean, which is obvious from \eqref{m.n.A}. On the other hand, if the \gls{cmm} approach is used, this is true if the constellations $\mcA$ may be decomposed into constant modulus, zero-mean sub-constellations; the demonstration is easy and omitted for sake of space. For the most popular constellations such as $M$-\gls{qam}, both conditions hold.}  Then, the only useful information is obtained from the non-zero $z_{\tilde{\gamma},n}$ (\ie from the pilots).

Thus,  the \gls{mp} in \eqref{z.alpha.init}-\eqref{z.beta} may be then seen as a \emph{pilot-only} based recursive estimation of the phase.  In fact, it is independent of the form of the constellation $\mcA$ used to modulate the payload symbols and this is why, in order to exploit the knowledge of $\mcA$, the \gls{cbc} phase tracking must be placed in the ``decoding-loop''. Known also as the joint phase tracking and decoding, it consists in alternate execution of  i)~the \gls{mp}  \eqref{z.alpha.init}-\eqref{z.beta} and ii)~the demodulation/decoding; the latter, providing an informative prior $\hat{P}^{\tr{a}}_n(a)$ on the payload symbols, improves the phase tracking, see \eg \cite{Colavolpe05,Barbieri07,Alfredsson19}.

\subsection{Intra-recursion projection: SKR algorithm}\label{Sec:Intra-MP}

As we said, the disadvantage of the pre-\gls{mp} projection from \secref{Sec:Pre-MP} is that, without the decoder's feedback, the structure of the constellation $\mcA$ is ignored. In order to exploit the form of $\mcA$, Shayovitz and Raphaeli \cite{Shayovitz16}, as well as, Kreimer and Raphaeli \cite{Kreimer18} proposed to perform the projection on $\mcF$ \emph{after} the multiplications in \eqref{alpha}-\eqref{beta} are executed.  That is, the projections are carried out inside the \gls{mp} recursion.

From the perspective we adopted, the resulting \acrfull{skr} algorithm is obtained implementing  the approximations \eqref{alpha.projection.preMP.1}
\begin{align}
\alpha_n(\theta)&\propto 
\label{alpha.projection.3}
 \mcF\big[\alpha_{n-1}(\theta)\gamma_{n-1}(\theta)\big]\convop \omega(\theta)\\
&\approx \check{\alpha}_{n-1}(\theta) \convop \omega(\theta),\\
\label{beta.projection}
\beta_n(\theta)&\propto 
\mcF\big[\beta_{n+1}(\theta)\gamma_{n+1}(\theta)\big]  \convop  \omega(\theta)\\
&\approx\check{\beta}_{n+1}(\theta)\convop  \omega(\theta),
\end{align} 
where we define the ``auxiliary'' distributions obtained via projections
\begin{align}
\nonumber
\check{\alpha}_{n}(\theta)&=\mcT\big(\theta; z_{\check{\alpha},n}\big)\\
\label{check.alpha}
&=\mcF\left[\sum_{a\in\mcA} P^{\tr{a}}_{n}(a) \alpha_{n}(\theta)g_{n}(\theta, a )\right]\\
\label{z.alpha.check}
z_{\check{\alpha},n}&=\mf{CMM}\Big[\big\{\eta_{g,n}(a), z_{\alpha,n}+z_{g,n}(a)\big\}_{a\in\mcA} \Big].
\end{align}
and
\begin{align}
\nonumber
\check{\beta}_{n}(\theta)&=\mcT\big(\theta; z_{\check{\beta},n}\big)\\
\label{check.beta}
&=\mcF\left[ \sum_{a\in\mcA} P^{\tr{a}}_{n}(a) \beta_{n}(\theta)g_{n}(\theta, a )  \right],\\
\label{z.beta.check}
z_{\check{\beta},n}&=\mf{CMM}\Big[\big\{\eta_{g,n}(a), z_{\beta,n}+z_{g,n}(a)\big\}_{a\in\mcA} \Big].
\end{align}

Using \eqref{Tik.h.Normal.convolution} in \eqref{alpha.projection.3} and \eqref{beta.projection} we obtain the following parametric \gls{mp}:
\begin{align}
\label{z.alpha.0.SKR}
z_{\alpha,0}&=0,\quad z_{\check\alpha,0}=z_{g,0}(x_0)\\
\label{z.alpha.n.SKR}
z_{\alpha,n}&= \frac{z_{\check{\alpha},n-1}}{1+|z_{\check{\alpha},n-1}| \sigma_w^2}, \quad n=1, \ld,N-1\\
\label{z.check.alpha.n.SKR}
z_{\check{\alpha},n}&=\mf{CMM}\Big[\big\{\eta_{g,n}(a), z_{\alpha,n}+z_{g,n}(a)\big\}_{a\in\mcA} \Big]\\
\label{z.beta.0.SKR}
z_{\beta,N}&=0,\quad z_{\check\beta,N}=z_{g,N}(x_N)\\
\label{z.beta.n.SKR}
z_{\beta,n}&= \frac{z_{\check{\beta},n+1}}{1+|z_{\check{\beta},n+1}| \sigma_w^2}, \quad n=N-1,\ld,1\\
\label{z.check.beta.n.SKR}
z_{\check{\beta},n}&=\mf{CMM}\Big[\big\{\eta_{g,n}(a), z_{\beta,n}+z_{g,n}(a)\big\}_{a\in\mcA} \Big].
\end{align}

Comparing to the  \gls{cbc} algorithm from \secref{Sec:Pre-MP}, the \gls{skr} algorithms invokes the \gls{cmm} function twice for each time $n$ and this, even if no feedback is obtained from the decoder. This increases the computational burden comparing to the \gls{cbc} algorithm, which  invokes the \gls{cmm} function only once per $n$ (and does not need it at all in the case the decoder's feedback is not available). The hope is that, introducing approximations (projection) inside the \gls{mp} recursion will minimize the approximation errors.

Again, we can apply the \gls{skr} algorithm iteratively, using the decoder's feedback which is taken into account via $P^{\tr{a}}_n(a)$.

\subsection{SKR and \cite{Shayovitz16,Kreimer18}}\label{Sec:Intra.vs.Raphaeli}

The \gls{skr} algorithm should be seen as a ``canonical'' representation of the algorithms shown in \cite{Shayovitz16,Kreimer18}: it relies on the Tikhonov distribution as suggested by \cite{Shayovitz16}, while the projections defined in \eqref{alpha.projection.3} and \eqref{beta.projection} are the essence of the algorithm proposed in \cite{Kreimer18}.


On the other hand, the choice of the family $\mcF$ in \cite{Kreimer18} and in \cite{Shayovitz16} leads to a certain ambiguity of implementation of the algorithms. In particular, 
\begin{itemize}
\item In \cite{Shayovitz16}: because $\mcF$ which is a \emph{mixture} of  Tikhonov distributions, the projection requires clustering of the elements of the mixture. This is computationally complex procedure even for moderate-size problems.\footnote{This is because even if the order of the reduced mixture is fixed to $L<M$ clusters, there is a formidably large number of possible assignments of $M$ elements into $L$ groups and it is usually unfeasible to test all possibilities} The heuristics are thus required and depend very much on the implementation details (such as, for example, enumeration order of the elements of the mixture or the thresholds used in the clustering). Further, despite the heuristics, a non-negligible complexity is required by the algorithms proposed in \cite{Shayovitz16}. In the follow-up work \cite{Kreimer18} (by the same author as in \cite{Shayovitz16}) the mixture was replaced by single (Gaussian) distribution.\footnote{In fact, \cite{Kreimer18} says about  \cite{Shayovitz16}  ``However, good results for high order constellations require much higher complexity and were not demonstrated.'' This suggest that the excessive complexity is the very reason why the author of \cite{Kreimer18} refrained from showing the results of his own algorithm developed in \cite{Shayovitz16} even for $M=16$.}

\item In  \cite{Kreimer18}: using the Gaussian family $\mcF$, the approximations must start already when representing $g_n(\theta,a)$. In fact,  \cite{Kreimer18} describes in details different strategies for such approximations but there is no unique answer to how this should be done.
\end{itemize}

In fact, the \gls{skr} algorithm is a particular version of the algorithm shown in \cite{Shayovitz16} (obtained if we force the family $\mcF$ to contain one Tikhonov distribution, which may be seen as a ``degenerate'' mixture). 

However, implementing the \gls{skr} algorithm using the \gls{cmm} projection formulas derived in \cite[Appendix A]{Shayovitz16} leads to significant error, see \figref{fig:PER.SKR+EP.OneShot}. This is because the last approximation step, expressed in \cite[Eq.~(102)]{Shayovitz16},  assumes that all terms of the mixture have a similar circular mean and  this assumption does not materialize in practice. On the other hand, implementing the approximation step defined in \cite[Eq.~(101)]{Shayovitz16} yields the results similar to those obtained by the \gls{skr} algorithm we defined in this work. 

This  observation is anecdotical but confirms the importance of  testing the approximations in the projection formulas; this is the goal of \appref{Sec:CMM} where the derivation are approximations-free.

\section{Expectation Propagation: Self-iterations in the phase-tracking}\label{Sec:It.EP}

As we will see in \secref{example.start}, the performance of the one-shot \gls{dmp} indicates that satisfactory solutions may be obtained without relying on the decoders' feedback. Therefore, our goal will be to develop a  reliable parametric one-shot phase tracking by exploiting the form of the constellation $\mcA$.

To explain why this can be done we note that, although the phase tracking \gls{mp} is defined on the tree (graph) that is known to yield the optimal solution in one run (backward-forward processing),  this is only  true for the \gls{dmp} which tracks the entire (discretized) distributions. On the other hand, the \gls{mp} in the \gls{cbc} and the \gls{skr} algorithm is based on approximated distributions $\tilde{\gamma}_n(\theta)$ (in \gls{cbc}) or $\check{\alpha}_n(\theta)$ and $\check{\beta}_n(\theta)$ (in \gls{skr}), therefore, a guarantee of optimality does not exist.

To remedy the resulting sub-optimality, we will improve the approximations iteratively; this will be done with ``self-iterations", that is, without any help from the decoder: the priors $\hat{P}^\tr{a}(a)$ will not change during the self-iterations. 

We follow here the \gls{ep} idea \cite[Ch.~3.2]{Minka01} which addresses the very problem of using approximate distributions in the \gls{mp}. We will apply the projection $\mcF[\cd]$ many times (iteratively) taking into account the results obtained in previous iterations.  Since most of the resulting operations are almost identical to those we already defined in the \gls{cbc} and the \gls{skr} algorithms, we will use the parenthesized superscript $^{(i)}$ to denote the variables/functions obtained in the $i$-th iteration, where $i=1,\ld, I_\tr{ep}$ and $I_\tr{ep}$ is the maximum number of \gls{ep} iterations.

\subsection{Expectation propagation in CBC algorithm}\label{Sec:EP.CBC}


The \gls{ep} applied to our problem relies on the following idea: in the iteration $i$,  to find the approximation $\tilde{\gamma}^{(i)}_n(\theta)$ we will rely on the distributions $\alpha^{(i-1)}_n(\theta)$ and $ \beta^{(i-1)}_n(\theta)$ calculated in the previous iteration. This is done by exploiting the posterior distributions of $\theta_n$ after the iteration $i-1$ which is a mixture given by
\begin{align}
\label{posterior.i-1}
f^{(i-1)}_n(\theta)\propto \gamma_n(\theta)\big(\alpha^{(i-1)}_n(\theta)\beta^{(i-1)}_n(\theta)\big).
\end{align}

Our objective is to find $\tilde{\gamma}^{(i)}_n(\theta)\in\mcF$ which generates the approximate posterior
\begin{align}
\label{posterior.i.i-1}
\tilde{f}^{(i-1,i)}_n(\theta)\propto \tilde\gamma^{(i)}_n(\theta)\big(\alpha^{(i-1)}_n(\theta)\beta^{(i-1)}_n(\theta)\big)
\end{align}
close to $f^{(i-1)}_n(\theta)$, that is, we require
\begin{align}
\nonumber
\tilde{\gamma}^{(i)}_n(\theta)\big(\alpha^{(i-1)}_n(\theta)&\beta^{(i-1)}_n(\theta)\big)\\
\label{equal.proba.approx}
&\approx
\gamma_n(\theta)\big(\alpha^{(i-1)}_n(\theta)\beta^{(i-1)}_n(\theta)\big).
\end{align}

To satisfy \eqref{equal.proba.approx} we apply the projection, $\mcF[\cd]$ to its both sides. Since the \gls{lhs} contains the three terms from the family  $\mcF$, it is not affected by the projection (remember, the product of the Tikhonov distributions is a also a Tikhonov distribution), we obtain
\begin{align}\label{equal.proba.approx.2}
\tilde{\gamma}^{(i)}_n(\theta)
&\propto
\frac{\mcF\Big[\gamma_n(\theta)\big(\alpha^{(i-1)}_n(\theta)\beta^{(i-1)}_n(\theta)\big)\Big]}
{\alpha^{(i-1)}_n(\theta)\beta^{(i-1)}_n(\theta)}.
\end{align}

Of course, for $i=1$, we deal with non-informative distributions $\alpha^{(0)}(\theta)\propto 1$ and $\beta^{(0)}(\theta)\propto 1$, which reduces \eqref{equal.proba.approx.2} to  $\tilde{\gamma}^{(1)}_n(\theta)=\mcF\big[ \gamma_n(\theta)\big]$; this is what was done in \secref{Sec:Parametric.MP}, \ie  $\tilde{\gamma}_n^{(1)}(\theta)$ is the same as $\tilde{\gamma}_n(\theta)$ obtained in \eqref{gamma.projection.first}.\footnote{Similarly, if we want to apply the projection operator, $\mcF[\cd]$, to each of the terms under multiplication in the numerator in \eqref{equal.proba.approx.2}, the result will be the same as the original \gls{cbc} algorithm: the projection  $\mcF\Big[\alpha^{(i-1)}_n(\theta)\beta^{(i-1)}_n(\theta)\Big]=\alpha^{(i-1)}_n(\theta)\beta^{(i-1)}_n(\theta)$ will cancel out with the denominator and iterative improvement will not be possible. This ``transparency" to the presence of $\alpha_n(\theta)$ and $\beta_n(\theta)$ is thus nothing but the conventional projection characteristic of the \gls{cbc} algorithm and, although it may provide satisfactory results in other context, \eg \cite{Vannucci20}, here it is not useful.}

For $i>1$, the informative distributions, $\alpha^{(i-1)}_n(\theta)$ and $\beta^{(i-1)}_n(\theta)$ are available so we have to calculate the \gls{rhs} of \eqref{equal.proba.approx}
\begin{align}
\gamma_n(\theta)\Big(\alpha^{(i-1)}_n(\theta)\beta^{(i-1)}_n(\theta)\Big)
\label{apost.theta.i1}
&=
\sum_{a\in\mcA} \e^{\eta_{g,n}(a)} \mcT\big(\theta; z^{(i-1)}_{\Sigma,n}(a) \big),
\end{align}
where 
\begin{align}
z_{\Sigma,n}^{(i-1)}(a)=z_{\alpha,n}^{(i-1)}+z_{\beta,n}^{(i-1)}+z_{g,n}(a)
\end{align}
and we see that  $z_{\Sigma,n}^{(1)}(a)$ is the same as \eqref{z.Sigma.n}.

Using \eqref{apost.theta.i1} in the numerator of \eqref{equal.proba.approx.2} yields
\begin{align}\label{EP.iterations}
\tilde{\gamma}^{(i)}_n(\theta) 
&=\mcT(\theta; z^{(i)}_{\tilde{\gamma},n})\\
\label{EP.division}
&\propto \frac
{\mcF\Big[\sum_{a\in\mcA} \e^{\eta_{g,n} (a)} \mcT\big(\theta; z^{(i-1)}_{\Sigma,n}(a) \big)\Big] }
{\mcT\big(\theta; z^{(i-1)}_{\alpha,n}+z^{(i-1)}_{\beta,n} \big)}\\
\label{EP.division.2}
&= \frac{ \mcT\Big(\theta; z^{(i)}_{\tr{post},n}\Big)  }{ \mcT\Big(\theta; z^{(i-1)}_{\alpha,n}+z^{(i-1)}_{\beta,n} \Big)},
\end{align}
where 
\begin{align}
z^{(i)}_{\tr{post},n} &=\mf{CMM}\Big[ \set{\eta_{g,n}(a), z^{(i-1)}_{\Sigma,n}(a)}_{a\in\mcA}  \Big]
\end{align}
and thus
\begin{align}
\label{z.gamma.i}
z^{(i)}_{\tilde{\gamma},n}&= z^{(i)}_{\tr{post},n}  - \big(z^{(i-1)}_{\alpha,n} + z^{(i-1)}_{\beta,n}\big).
\end{align}

\subsubsection{Pitfalls  and solutions}\label{Sec:Pitfalls}
The elegant formulation of  the \gls{ep} is indeed appealing but occasionally runs into difficulty due to the division of the distributions in \eqref{EP.division.2}, or equivalently, due to subtraction of the coefficients in \eqref{z.gamma.i}. The existence of a similar problem has been acknowledged in \cite[Sec.~3.2]{Minka01} in the context of the Gaussian \gls{ep}, where the division of the distributions may yield a ``Gaussian'' with negative variance, see also \cite[Sec.~IV.B]{Senst11} \cite[Sec.~IV.A]{Cespedes14}. Such results are usually uninterpretable so a workaround is needed. 

In the context of the \gls{ep} based on Tikhonov distributions $\mcT(\theta;z)$ which are defined by complex parameters $z$, the purely numerical issue of invalid parameters (such as a negative variance in  the Gaussian case) is avoided but the problem remains. To understand intuitively its source and devise a solution, we may look at the scenario where the problems materialize. 

\begin{example}[Pitfalls of EP]\label{Ex:Pitfall}
Assume that 
\begin{itemize}
\item The circular means of the approximate posterior distribution, $\mcT\big(\theta; z^{(i)}_{\tr{post},n}\big) $ and of the ``extrinsic'' distribution (defined by  the product $\alpha^{(i-1)}_n(\theta)\beta^{(i-1)}_n(\theta)$) coincide, that is,  $\angle z^{(i)}_{\tr{post},n} =\angle (z^{(i-1)}_{\alpha,n}+z^{(i-1)}_{\beta,n})$; (remember, the phase $\angle z^{(i)}_{\tr{post},n}$ defines the circular mean); and that
\item The variance of the posterior distribution is larger than the variance of the extrinsic distribution, \ie $|z^{(i)}_{\tr{post},n}|<|z^{(i-1)}_{\alpha,n}+z^{(i-1)}_{\beta,n}|$; this may occur because the distribution in the argument of $\mcF[\cd]$ in \eqref{EP.division} is obtained multiplying  the extrinsic distribution $\mcT\big(\theta; z^{(i-1)}_{\alpha,n}+z^{(i-1)}_{\beta,n}\big)$ with a mixture $\gamma_n(\theta)$; the variance of the resulting mixture (and thus of its projection result as well) may be larger than the variance of the extrinsic distribution. 
\end{itemize}

Then, carrying out the subtraction in \eqref{z.gamma.i} we will obtain
\begin{align}
z^{(i)}_{\tilde{\gamma},n}
&= (|z^{(i)}_{\tr{post},n}|  - |z^{(i-1)}_{\alpha,n} + z^{(i-1)}_{\beta,n}|\big)\e^{\jj \angle z^{(i)}_{\tr{post},n}}\\
&=|z^{(i-1)}_{\tilde{\gamma},n}|\e^{\jj(\angle z^{(i)}_{\tr{post},n}+\pi)}.
\end{align}
That is, the circular mean of the new distribution $\mcT(\theta; z^{(i)}_{\tilde{\gamma},n})$ will be  in disagreement (by the largest possible value of $\pi$) with the mean of the posterior and extrinsic distributions obtained from the previous iteration. Such a results, being an artefact of the way the \gls{ep} is defined \cite[Sec.~3.2]{Minka01} is clearly counterintuitive and simply wrong. 
\end{example}

This problem, characteristic of the \gls{ep} will affect first $\tilde{\gamma}_n(\theta)$ and,  propagating via the \gls{mp}, will have detrimental affects on $\alpha_n(\theta)$ and $\beta_n(\theta)$.

Heuristics were devised for the Gaussian \gls{ep} where the problem is clearly identified, \ie the variance of the distributions after division becomes negative. For example, \cite[Sec.~3.2]{Minka01} constrains the variance to be positive, while other proposed to identify the problematic cases (\ie the negative variance) and then, if necessary i)~eliminate the division of the distributions \cite[Sec.~IV.B]{Senst11}, which here would mean $z^{(i)}_{\tilde{\gamma},n}=z^{(i)}_{\tr{post},n}$, or ii)~remove the update \cite[Sec.~IV.A]{Cespedes14}, \ie $z^{(i)}_{\tilde{\gamma},n}=z^{(i-1)}_{\tilde{\gamma},n}$. 

On the other hand, instead of testing for the compliance with our prior requirements (that are not necessarily obvious to define),  we might ``smooth'' the obtained parameters via recursive filter \cite[Sec.~IV.A]{Cespedes14} , which would mean  replacing \eqref{z.gamma.i} with
\begin{align}\nonumber
z_{\tilde{\gamma},n}^{(i)} = \zeta \Big(z^{(i)}_{\tr{post},n}  - z^{(i-1)}_{\alpha,n} - z^{(i-1)}_{\beta,n}\Big)
&+(1-\zeta) z_{\tilde{\gamma},n}^{(i-1)}\\
\label{z.smoothing}
& i=2,\ld, I_\tr{ep},
\end{align}
where the smoothing parameter, $\zeta<1$, must be chosen heuristically to strike a balance between the new solution and the history accumulated in the previous estimate. In this way we avoid rapid changes in the the parameters $z_{\tilde{\gamma},n}^{(i)}$, regularizing the final solution. Note that the smoothing is not necessary in the first iteration, $i=1$, which is based solely on the information obtained from the pilots (and from the decoder if we allows for it). 


\subsubsection{Summary of the algorithm}\label{Sec:Summary}
The proposed \gls{cbc}+\gls{ep} phase tracking algorithm is now summarized as Algorithm~\ref{Algo:EP} where we integrated the notation so that the \gls{cbc} algorithm is naturally the first iteration ($i=1$) of the \gls{cbc}+\gls{ep} algorithm, \ie $z^{(0)}_{\Sigma,n}(a)=z_{g,n}(a)$ . This first iteration is also distinct: the smoothing \eqref{z.smoothing} is not applied for $i=1$, see lines \ref{i.if}--\ref{i.endif} of the algorithm. Note that the notation with the iteration index $^{(i)}$ may be removed (and the in-place calculation carried out) but we kept it for compatibility with the equations in the paper.

Last but not least, and at the risk of stating the obvious, we want to emphasize what the \gls{cbc}+\gls{ep} algorithm is \emph{not} doing. Namely, it is not reusing the output ``extrinsic'' log-probabilities $\hat{P}_n(a)$ as if they were newly calculated prior log-probabilities $\hat{P}^{\tr{a}}_n(a)$. This is clearly seen in the description of the algorithm: the prior probability $\hat{P}^{\tr{a}}_n(a)$ affects the weighting factors $\eta_{g,n}(a)$ in line \ref{eta.gn.a} which remain unaltered throughout the \gls{cbc}+\gls{ep} iterations; the \gls{cbc}+\gls{ep} algorithm rather changes the parameters of the Tikhnov distributions $z^{(i)}_{\Sigma,n}(a)$.

\begin{algorithm}
    \caption{CBC+EP phase tracking} \label{Algo:EP}
\begin{algorithmic}[1]
\State Inputs:
\State $y_n, n\in\set{0,\ld,N}$ \Comment Received signal 
\State $x_n, n\in\Natural_{\tr{pilots}}$  \Comment Pilot symbols
\State $\hat{P}^{\tr{a}}_n(a)$\Comment Log-probabilities from the decoder
\State Initialization:
\State $z^{(i)}_{\tilde{\gamma},n}\gets 2\SNRrv~ y_n x_n^*,   \quad  n\in \Natural_{\tr{pilots}}, i\in{1,\ld, I_\tr{ep}}$\label{z.i.tilde.pilot}
\State $\eta_{g,n}(a) \gets -\SNRrv~|a|^2+\hat{P}^{\tr{a}}_n(a),\quad a\in\mcA, n\in\Natural_{\tr{payload}}$\label{eta.gn.a}
\State $z_{g,n}(a)\gets 2\SNRrv~y_n a^*,\quad a\in\mcA, n\in\Natural_{\tr{payload}}$\label{z.g.n.payload}
\State $z_{\Sigma,n}^{(0)}(a) \gets z_{g,n}(a), \quad n\in\Natural_{\tr{payload}}$
\State EP iterations:
\For{ $i\gets1,\ld, I_\tr{ep}$ }
	\State Pre-MP : Finds $\tilde{\gamma}^{(i)}_n(\theta)$
	\For {$n\gets \Natural_{\tr{payload}}$} 
		\State $z^{(i)}_{\tr{post},n}  \gets \mf{CMM}\Big[ \eta_{g,n}(a), \set{z^{(i-1)}_{\Sigma,n}(a)}_{a\in\mcA}  \Big]$
	\If {$i=1$} \label{i.if}
		\State $z^{(i)}_{\tilde{\gamma},n}\gets z^{(i)}_{\tr{post},n}$
	\Else
		\State $z_{\tilde{\gamma},n}^{(i)} \gets \zeta \Big(z^{(i)}_{\tr{post},n}  - z^{(i-1)}_{\alpha,n} - z^{(i-1)}_{\beta,n}\Big)$
		\State \qquad~ \qquad  ~ \qquad~ \qquad~ \qquad$+(1-\zeta) z_{\tilde{\gamma},n}^{(i-1)}$
	\EndIf  \label{i.endif}
	\EndFor
	\State MP recursive calculation:
	\State $z^{(i)}_{\alpha,0}\gets 0$	
	\For {$n\gets 1,\ld,N$} \Comment Finds $\alpha^{(i)}_n(\theta)$:
		\State $z^{(i)}_{\alpha,n} \gets \frac{z^{(i)}_{\alpha,n-1} + z^{(i)}_{\tilde{\gamma},n-1}}{1+\big|z^{(i)}_{\alpha,n-1} + z^{(i)}_{\tilde{\gamma},n-1}\big|\sigma_w^2}$
	\EndFor
   	\State $z^{(i)}_{\beta,N} \gets 0$
	\For {$n\gets N-1,\ld,0$} \Comment Finds $\beta^{(i)}_n(\theta)$:
		\State $z^{(i)}_{\beta,n} \gets  \frac{z^{(i)}_{\beta,n+1} + z^{(i)}_{\tilde{\gamma},n+1}}{1+\big|z^{(i)}_{\beta,n+1} + z^{(i)}_{\tilde{\gamma},n+1}\big|\sigma_w^2}$
	\EndFor
	\For {$n\gets \Natural_{\tr{payload}}$} 
	\State $z^{(i)}_{\Sigma,n}(a) \gets  z_{g,n}(a) +z^{(i)}_{\alpha,n}+ z^{(i)}_{\beta,n}$
	\EndFor
\EndFor 
\State Output: log-probabilities to be used in \eqref{lambda.n.l}
\For {$n \gets \Natural_{\tr{payload}}$}
\State $\hat{P}_{n}( a ) \gets -\SNRrv |a|^2 + \hat{I}_0\big(|z^{(I_{\tr{ep}})}_{\Sigma,n}(a)|\big)$\label{P.n.a}
\EndFor
\end{algorithmic}
\end{algorithm}

\subsection{Expectation propagation in SKR algorithm}\label{Sec:EP.SKR}

The reasoning behind the \gls{ep} we used to obtain the \gls{cbc}+\gls{ep} algorithm will be now applied to enhance the \gls{skr} algorithm. Armed with the previous considerations, we already know how to proceed: in the iteration $i$ we first project on $\mcF$ the mixture which represents the posterior distribution which is calculated using results obtained in the iteration $i-1$. The latter are next removed from the projection results. 

The difference with the previous section is that we are not seeking to improve $\gamma_n(\theta)$ as we did in the case of the \gls{cbc} algorithm, but rather we want to obtain improved estimates of $\check\alpha_n(\theta)$ and $\check\beta_n(\theta)$. Thus, we have to transform \eqref{check.alpha}-\eqref{z.beta.check} as follows:
\begin{align}
\label{check.alpha.i}
\check{\alpha}^{(i)}_{n}(\theta)
&=\frac{\mcF\left[\sum_{a\in\mcA} P^{\tr{a}}_{n}(a) \alpha^{(i)}_{n}(\theta)g_{n}(\theta, a )\beta^{(i-1)}_n(\theta)\right]}{\beta^{(i-1)}_n(\theta)}
\end{align}
and
\begin{align}
\label{check.beta.i}
\check{\beta}^{(i)}_{n}(\theta)
&=\frac{\mcF\left[ \sum_{a\in\mcA} P^{\tr{a}}_{n}(a) \beta^{(i)}_{n}(\theta)g_{n}(\theta, a )  \alpha^{(i-1)}_n(\theta)\right]}{\alpha^{(i-1)}_n(\theta)},
\end{align}
where again we used indexing with $^{(i)}$ to denote the results from the iteration $i$.

The resulting SKR+EP algorithm is obtained applying this iteration-indexing in \eqref{z.alpha.0.SKR}-\eqref{z.check.beta.n.SKR} and using \eqref{check.alpha.i}-\eqref{check.beta.i}
\begin{align}
\label{z.alpha.0.SKR.i}
z^{(i)}_{\alpha,0}&=0,\quad z^{(i)}_{\check\alpha,0}=z_{g,0}(x_0),\\
\label{z.alpha.n.SKR.i}
z^{(i)}_{\alpha,n}&= \zeta \frac{z^{(i)}_{\check{\alpha},n-1}}{1+|z^{(i)}_{\check{\alpha},n-1}| \sigma_w^2}
    +(1-\zeta)z^{(i-1)}_{\alpha,n}, \\
\nonumber
z^{(i)}_{\check{\alpha},n}&=\mf{CMM}\Big[\big\{\eta_{g,n}(a), z^{(i)}_{\alpha,n}+z_{g,n}(a)+z_{\beta,n}^{(i-1)}\big\}_{a\in\mcA} \Big], \\
\label{z.alpha.check.i}
   & \qquad ~ \qquad - z_{\beta,n}^{(i-1)},\\
\label{z.beta.0.SKR.i}
z^{(i)}_{\beta,N}&=0,\quad z^{(i)}_{\check\beta,N}=z_{g,N}(x_N),\\
\label{z.beta.n.SKR.i}
z^{(i)}_{\beta,n}&= \zeta \frac{z^{(i)}_{\check{\beta},n+1}}{1+|z^{(i)}_{\check{\beta},n+1}| \sigma_w^2}+(1-\zeta)z^{(i-1)}_{\beta,n},\\
\nonumber
z_{\check{\beta},n}^{(i)}&=\mf{CMM}\Big[\big\{\eta_{g,n}(a), z^{(i)}_{\beta,n}+z_{g,n}(a) + z_{\alpha,n}^{(i-1)}\big\}_{a\in\mcA} \Big]\\
\label{z.beta.check.i}
   & \qquad ~ \qquad - z_{\alpha,n}^{(i-1)},
\end{align}
where, the division by $\beta^{(i-1)}_n(\theta)$ and by $\alpha^{(i-1)}_n(\theta)$ appearing, respectively in \eqref{check.alpha.i} and \eqref{check.beta.i} translates into subtraction of $z_{\beta,n}^{(i-1)}$ and of $z_{\alpha,n}^{(i-1)}$ given, respectively by \eqref{z.alpha.check.i} and  \eqref{z.beta.check.i}. 

In \eqref{z.alpha.n.SKR.i} and \eqref{z.beta.n.SKR.i} we also introduce the smoothing via weighting with $\zeta$ akin to the operation shown in \eqref{z.smoothing}. For the sake of space we refrain from explicating all operations as we did for the \gls{cbc}+\gls{ep} algorithm.

\subsection{Hybrid algorithm: CBC+SKR}\label{Sec:Hybrids}

We note that both algorithms, \gls{cbc}+\gls{ep} as well as \gls{skr}+\gls{ep}, calculate in each iteration the distributions $\alpha^{(i)}_n(\theta)$ and $\beta^{(i)}_n(\theta)$  which are then used in the next  iteration according to the corresponding formulas. However, in the iteration $i$, we place no restriction on \emph{how} the distributions $\alpha^{(i-1)}_n(\theta)$ or $\beta^{(i-1)}_n(\theta)$ were obtained. This suggest that we might reuse the results obtained from different algorithms. 

In particular, we might carry out the first iteration via the \gls{cbc} algorithm and then, in the second iteration use the obtained estimates of $\alpha^{(1)}_n(\theta)$ or $\beta^{(1)}_n(\theta)$ in the SKR algorithm. We denote such a hybrid approach by CBC+SKR. 

Noting that the \gls{cbc} algorithm is merely the pilot-based phase estimation, the CBC+SKR algorithm provides an interesting insight into the formal use of pilots  in two-stage phase estimation.

In fact, a similar two-stage approach can be found in the literature. For example, in \cite{Kreimer18}, the second stage consists of a regular parametric \gls{mp} implemented using a Gaussian family $\mcF$. But, in order to approximate $g_n(\theta,a)$ with a Gaussian distribution, we need to know which values of $\theta$ are the most relevant. This is done in the first stage: pilots are used to obtain ``raw" estimates of the phase $\hat{\theta}_n$, and the approximations are found for $\theta\approx\hat{\theta}_n$. 

Thus, in \cite{Kreimer18}, the pilot-based phase estimation (first stage) is related to the approximation problems rather than to the \gls{mp} formulation itself (but, of course, better approximation improves the results of the \gls{mp}). This is different from the CBC+SKR algorithm where we are not concerned at all with approximation issues but rather exploit the probabilistic model of the pilot-based phase estimates (obtained by the CBC algorithm) to modify the \gls{mp} in the \gls{skr} algorithm.

We show an example of the results obtained using  CBC+SKR algorithm in \secref{example.start}.

\subsection{Comments on complexity}\label{Sec:Sch.Cplx}

The complexity of decoding and demodulation (\glspl{llr} calculation) is considered to be common in all algorithms so we are concerned only with the complexity of the phase tracking algorithms per se, where the main complexity resides in the \gls{cmm} and more precisely in the calculation of the circular moment as defined by \eqref{mu.g.CMM.def} and the operation \eqref{z.tilde.def.CMM}: we need to carry out $M$ non-linear operations (function $B(x)/x$), $M$ absolute value calculations, $M$ multiplications, $M$ complex multiplication and $M$ additions. 

Thus the ``\gls{cmm} complexity'' is a unit may be the basis for comparison. 

The \gls{cbc} algorithm is notably simple because we do not need the \gls{cmm} at all; remember, the projection without decoder's feedback yields $z_{\tilde{\gamma},n}=0$ (for payload) and $z_{\tilde{\gamma},n}=z_{g,n}(a)$ (for pilots). So its \gls{cmm} complexity is zero.

The algorithm CBC+EP($I_\tr{ep}$) has the complexity of $I_\tr{ep}-1$ (of CMM complexity) units, the SKR+EP($I_\tr{ep}$) algorithm has the complexity equal to $2I_\tr{ep}$ units, while  the hybrid algorithm CBC+SKR displays the complexity of two (2) units.

In this perspective, the hybrid algorithm, CBC+SKR, has the same complexity as the algorithm SKR. This is not exactly true because we need one addition (\eg to calculate $z^{(i)}_{\alpha,n}+z^{(i-1)}_{\beta,n}$ before running the \gls{cmm} in \eqref{z.alpha.check.i}) and another one (\eg to  subtract  $z^{(i-1)}_{\beta,n}$ after running the \gls{cmm} in \eqref{z.alpha.check.i}), but this two additions can be neglected comparing with the unit cost of the \gls{cmm}.

\section{Numerical Examples}\label{example.start}

All examples use $M$-\gls{qam} constellation with Gray mapping \cite[Sec.~2.5.2]{Szczecinski_Alvarado_book}; the proprietary \gls{ldpc} encoder with the rate $r=\frac{7}{8}$ produces the block of $\Nc=4032$ bits.  We consider two modulation/phase-noise setups following \cite[Fig.~8]{Kreimer18} i) 16-\gls{qam} with $\sigma_w=0.1~\tr{rad}\approx 5.7~\tr{deg}$, and ii) 64-\gls{qam} with $\sigma_w=0.05~\tr{rad}\approx 2.9~\tr{deg}$. The pilot symbols are pseudo-randomly drawn from a $4$-\gls{qam} constellation; the pilot spacing is set to $L\in\set{17,25}$ which allows us to place $L-1\in\set{16,24}$ payload symbols between pilots, see \eqref{payload.1}, so dummy symbols are not needed to fill the frame of transmitted symbols $\set{x_n}$.


The decoding is based on the min-sum algorithm with scaling of the check-nodes messages by a constant factor equal to $\rho=0.7$, see \cite{Xu14}; then, a fraction of dB loss was observed when comparing to the sum-product algorithm in the \gls{awgn} channel. The total number of decoding iterations equals $I_\tr{dec}=10$.

The \gls{dmp} is implemented using $N_\theta=64$ samples\footnote{\label{foot:DMP32}The numerical results obtained for $16$-\gls{qam} and  $N_\theta\in\set{32,64,128}$ were practically indistinguishable while a significant deterioration of the results was observed for $N_\theta=16$. This indicates that the heuristics  $N_\theta=8M$ used \eg in \cite{Peleg00,Colavolpe05,Kreimer18}, may be too conservative but this issue is out of scope here. } and its \gls{per} curve is the performance limit for all  phase tracking algorithms.

The ``AWGN'' curve corresponds to a hypothetical scenario when the phase-noise is removed and we only deal with the \gls{awgn}. Such results are unattainable no matter how sophisticated the phase-tracking and decoding strategy are, and we show them to illustrate the impact of the phase noise.

The \acrfull{per} is estimated after transmitting $10^5$ blocks or after occurence of $100$ blocks in error, whichever comes first. 

The demodulation/phase-tracking algorithms which use the decoder's feedback are referred to as ``iterative'' and those which don't -- as ``one-shot" algorithms.

\subsection{One-shot receivers}\label{Sec:One-shot}
We analyze first one-shot receivers which means that the \gls{mp} phase tracking algorithm is run only once and has no prior information about the symbols, \ie $\hat{P}^{\tr{a}}_n(a)\propto 0$. In the case of the \gls{cbc} algorithm, it means that the phase reference for all the symbols is obtained using only the pilots.

\subsubsection{CBC+EP: The smoothing parameter}
\begin{figure}[bt]
\begin{center}
\psfrag{DMPXXXX.XXXXX}[lc][lc][\siz]{DMP}
\psfrag{CBC}[lc][lc][\siz]{CBC}
\psfrag{AWGN}[lc][lc][\siz]{AWGN}
\psfrag{EP2-02}[lc][lc][\siz]{CBC+EP(2),$\zeta=0.2$}
\psfrag{EP2-04}[lc][lc][\siz]{CBC+EP(2),$\zeta=0.4$}
\psfrag{EP2-06}[lc][lc][\siz]{CBC+EP(2),$\zeta=0.6$}
\psfrag{EP2-08}[lc][lc][\siz]{CBC+EP(2),$\zeta=0.8$}
\psfrag{EP2-10}[lc][lc][\siz]{CBC+EP(2),$\zeta=1.0$}
\psfrag{ylabel}[lc][lc][\siz]{$\PER$}
\psfrag{xlabel}[lc][lc][\siz]{$\SNRrv$ [dB]}
\scalebox{\sizf}{\includegraphics[width=\sizfs\linewidth]{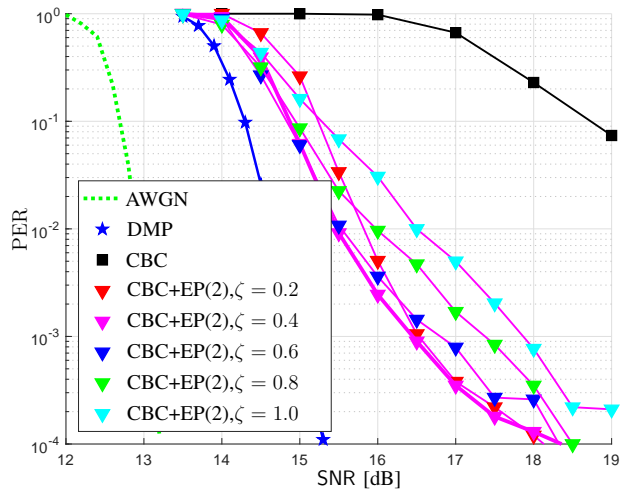}}
\end{center}
\caption{\gls{per} vs. \gls{snr} for one-shot algorithm in $16$-\gls{qam} transmission: CBC+\gls{ep}(2), \ie $I_\tr{ep}=2$, different values or the smoothing parameter $\zeta$. The thick lines indicate the results obtained for \mbox{$\zeta=0.4$} which is retained for further analysis. The results of the \gls{dmp} and \gls{cbc} algoritms are shown for reference.}\label{fig:PER.EP.OneShot}
\end{figure}

We start showing in \figref{fig:PER.EP.OneShot} the results produced by the CBC+EP($I_\tr{ep}$) algorithm for $I_\tr{ep}=2$ when using different values of the smoothing parameter $\zeta$, see \eqref{z.smoothing}. The best performance is obtained for $\zeta=0.4$ which will be used in all the remaining examples. Similar results were obtained for CBC+EP(3).

\begin{figure}[bt]
\begin{center}
\psfrag{DMPXXXX.XXXX}[lc][lc][\siz]{DMP}
\psfrag{CBC}[lc][lc][\siz]{CBC}
\psfrag{EP2 1.0}[lc][lc][\siz]{CBC+EP(2),$\zeta=1$}
\psfrag{EP2 0.4}[lc][lc][\siz]{CBC+EP(2),$\zeta=0.4$}
\psfrag{xlabel}[lc][lc][\siz]{$\theta$[deg]}
\psfrag{ylabel}[lc][lc][\siz]{$f_n(\theta)$}
\scalebox{\sizf}{\includegraphics[width=\sizfs\linewidth]{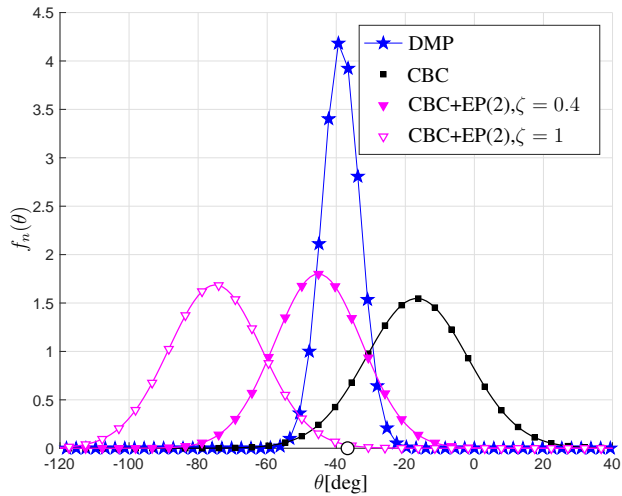}}
\end{center}
\caption{Example of the \gls{pdf}s obtained using one-shot CBC+EP(2) phase tracking algorithms; $L=25$, $\SNRrv=16$dB, $M=16$; the true value of the phase noise, $\theta_n\approx -38$deg, is indicated by the position of a circle on the axis of $\theta$.}\label{fig:PhaseSlip}
\end{figure}

These results confirm the existence of the problems we delineated in \secref{Sec:Pitfalls} and for illustration, we show in \figref{fig:PhaseSlip} an example of the \gls{pdf}s 
\begin{align}
\nonumber
f_n(\theta)&\propto \alpha_n(\theta)\beta_n(\theta)\\
&\propto f(\theta_n|y_0,\ld,y_{n-1},y_{n+1},\ld,y_N)
\end{align}
obtained using the \gls{dmp}, \gls{cbc} and CBC+\gls{ep}(2) algorithms. 

We also show the true value of the phase $\theta_n\approx-38\tr{deg}$ (a black circle on the axis) and we can see that while the \gls{dmp} algorithms provides a very reliable probabilistic estimation of the latter (its mode/mean being close to $\theta_n$), the \gls{cbc} algorithm, being based on pilots only is much less accurate. What is important here is to note that using the \gls{ep} algorithm with $\zeta=1$ yields the distribution $f_n(\theta)$ which also considerably misses the actual value of the phase: the difference between the circular mean $\angle(z^{(2)}_{\alpha,n}+z^{(2)}_{\beta,n})$ and $\theta_n$ is close to $40$deg. This is what \cite{Shayovitz16} calls a ``phase-slip''. 

We may also appreciate that, thanks to the smoothing operation in \eqref{z.smoothing}, the CBC+EP(2) algorithms with $\zeta=0.4$ mediates between  $\zeta=0$ (equivalent to the \gls{cbc} algorithm) and $\zeta=1$ (the second \gls{ep} iteration without smoothing). So, while the first iteration of the \gls{ep} algorithm being based on the pilots only is not very accurate, it is robust and may be used to regularize the results from the subsequent iterations: this is the role of smoothing  \eqref{z.smoothing}. 

Such  intuitive understanding of the observed phenomenon cannot entirely explain the mechanism behind this regularization which is rather involved due to the \gls{mp} recursion. Moreover, while choosing $\zeta=0.4$ improves the overall performance, there is no definite answer  which of the solutions ($\zeta=1$ or $\zeta=0$) is closer to the exact distribution (\gls{dmp}) for each $n$. Better  regularization strategies are  possible but this issue is a matter for separate investigation.

\subsubsection{SKR}\label{Sec:SKR.Numerical}

\begin{figure}[bt]
\begin{center}
\psfrag{DMPXXXX.XXXX}[lc][lc][\siz]{DMP}
\psfrag{SKR+EP2-04}[lc][lc][\siz]{SKR+EP(2),$\zeta=0.4$}
\psfrag{SKR+EP2-06}[lc][lc][\siz]{SKR+EP(2),$\zeta=0.6$}
\psfrag{SKR+EP2-08}[lc][lc][\siz]{SKR+EP(2),$\zeta=0.8$}
\psfrag{SKR+EP2-10}[lc][lc][\siz]{SKR+EP(2),$\zeta=1.0$}
\psfrag{SKR-Sh}[lc][lc][\siz]{SKR + \cite{Shayovitz16}}
\psfrag{SKR}[lc][lc][\siz]{SKR}
\psfrag{ylabel}[lc][lc][\siz]{$\PER$}
\psfrag{xlabel}[lc][lc][\siz]{$\SNRrv$ [dB]}
\scalebox{\sizf}{\includegraphics[width=\sizfs\linewidth]{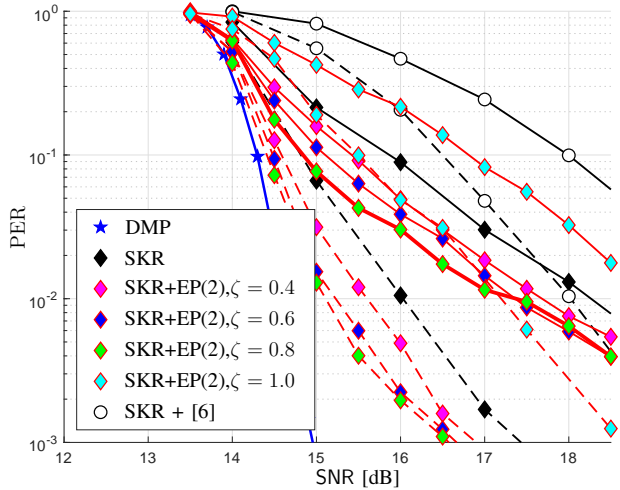}}
\end{center}
\caption{\gls{per} vs. \gls{snr} for one-shot algorithms in $16$-\gls{qam} transmission: DMP, SKR, SKR+EP(2) with different values or the smoothing parameter $\zeta$ (thick line for $\zeta=0.8$); $L=25$ (solid line) and $L=17$ (dashed lines). The results ``\gls{skr}+\cite{Shayovitz16}''  correspond to the \gls{skr} algorithm based on the \gls{cmm} defined by \cite[Eq.~(102)]{Shayovitz16}.}\label{fig:PER.SKR+EP.OneShot}
\end{figure}

Here, we repeat the experience from the previous section for the algorithm SKR+EP(2) and show the results in \figref{fig:PER.SKR+EP.OneShot}. Without smoothing (\ie with $\zeta=1$) the \gls{ep} deteriorates the results but even with appropriately chosen $\zeta=0.8$ the gains are much less notable than in the case of the CBC+EP algorithm and this, despite the complexity of four (4) CMM complexity units, see \secref{Sec:Sch.Cplx}. Quantitatively similar results are obtained for different pilot spacings,  $L=25$ and $L=17$. 

 We thus only show the results of the SKR algorithm in the remaining figures.

We also show the results obtained using the \gls{skr} algorithm but where the \gls{cmm} is implemented using the formulas from \cite[Appendix A]{Shayovitz16} (indicated by the ``SKR+\cite{Shayovitz16}'' curve). The poor performance is due to oversimplification of the formula \cite[Eq.~(102)]{Shayovitz16} which should be avoided. On the other hand, the approximation  \cite[Eq.~(101)]{Shayovitz16} produces results equivalent to the \gls{skr} algorithm we show. 

\subsubsection{CBC, SKR, and Hybrid algorithms}

\begin{figure}[bt]

\psfrag{DMPXXXX.XXXX}[lc][lc][\siz]{DMP}
\psfrag{CBC}[lc][lc][\siz]{CBC}
\psfrag{AWGN}[lc][lc][\siz]{AWGN}
\psfrag{Intra-MP}[lc][lc][\siz]{SKR}
\psfrag{Intra-Sh}[lc][lc][\siz]{SKR+\cite{Shayovitz16}}
\psfrag{EP2}[lc][lc][\siz]{CBC+EP(2)}
\psfrag{EP3}[lc][lc][\siz]{CBC+EP(3)}
\psfrag{HYBRID}[lc][lc][\siz]{CBC+SKR,$\zeta=0.8$}
\psfrag{ylabel}[lc][lc][\siz]{$\PER$}
\psfrag{xlabel}[lc][lc][\siz]{$\SNRrv$ [dB]}
\begin{tabular}{c}

\scalebox{\sizf}{\includegraphics[width=\sizfs\linewidth]{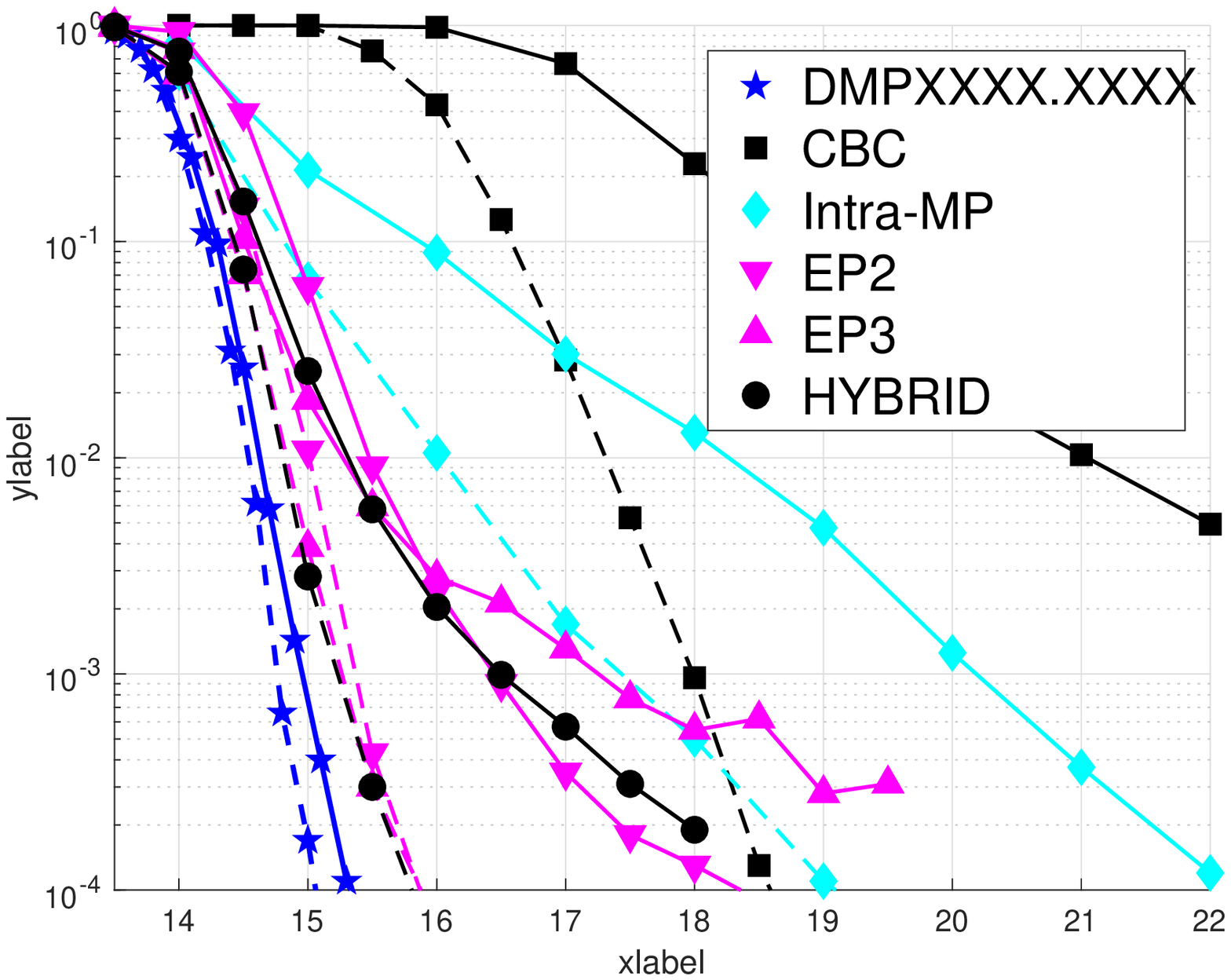}}\\
a)\\
~\\

\scalebox{\sizf}{\includegraphics[width=\sizfs\linewidth]{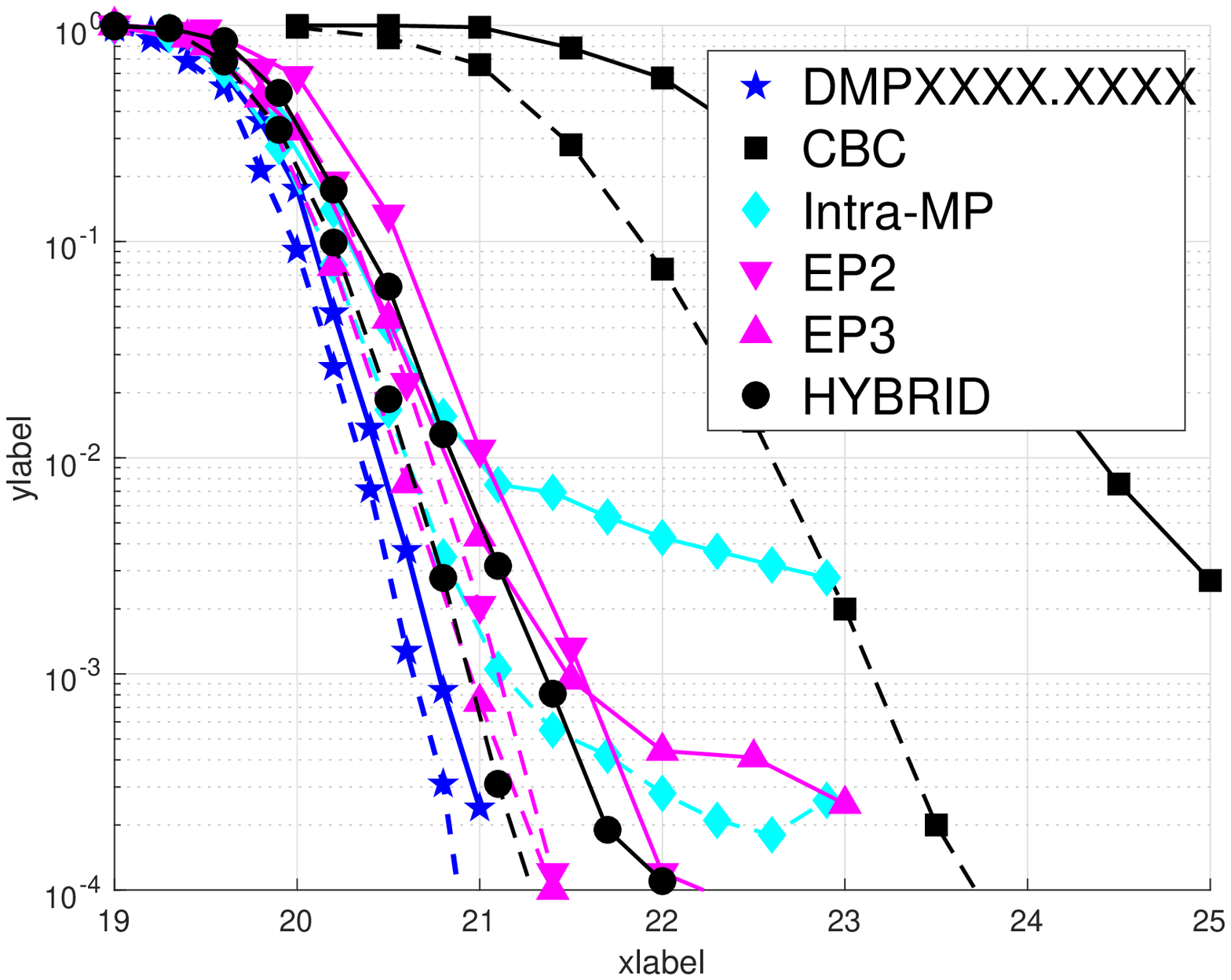}}\\
b)
\end{tabular}
\caption{\gls{per} vs. \gls{snr} for one-shot phase tracking algorithms in a) $16$-\gls{qam} and b) 64-\gls{qam} transmission; two pilot spacings $L=17$ (dashed) and $L=25$ (solid), are considered. The \gls{dmp} is based on $N_\theta=64$ samples.}\label{fig:PER.OneShot}
\end{figure}

The results obtained using the algorithms \gls{dmp}, \gls{cbc}, \gls{skr}, and CBC+SKR are now shown in \figref{fig:PER.OneShot} where we observe that 
\begin{enumerate}
\item The performance decreases notoriously with the increased pilot spacing for all the algorithm but for the \gls{dmp}, where the effect is much less notable. We may thus infer that it is indeed the projection on $\mcF$ which is the principal source of degradation of the parametric phase tracking algorithm.

\item The \gls{per} curves show a tendency to error floor  (\ie increasing the \gls{snr} yields small \gls{per} gains)\footnote{\label{Foot.DMP.Floor}Similar tendency is observed for the \gls{dmp} below $\PER=10^{-4}$ as will be shown later in \figref{fig:PER.Iterative.CBC}.} that is  especially notable for a large pilot spacing, $L=25$. This is because, for high \gls{snr}, the phase noise is a dominating distortion, however, the reasons behind error floor  are not exactly the same in all algorithms. In particular
\begin{itemize}
\item In one-shot \gls{cbc} algorithm, the mixture reduction for the payload symbols is trivial (\ie $z_{\tilde{\gamma},n}=0$) so the uncertainty about the phase depends only on the pilots. 
\item  The mixture reduction in the \gls{skr}, SKR+EP, and the CBC+\gls{ep} algorithms relies on non-trivial operations based on  the \gls{cmm}
which are prone to errors: the projection may over-emphasize one of the components $g_n(\theta,a)$ which does not represent the actual phase distribution (that is, when the component $g_n(\theta,a)$ dominating the mixture is the one with $a\neq x_n$); this leads to the phase-slip.
\end{itemize}

This algorithmically-produced phase-slip also explains why the \gls{skr} algorithm which, at each time $n$, exploits the form of the constellation $\mcA$, may be outperformed  by the \gls{cbc} algorithm (for $L=17$ in \figref{fig:PER.OneShot}) which is merely based on the pilots and ignores the form of $\mcA$: the respective \gls{per} curves cross for $\SNRrv\approx18.5$dB (when $M=16$) and $\SNRrv\approx23.5$dB (when $M=64$). 

In fact, this very dependence on the pilots \emph{only} is the source of the robustness of the \gls{cbc} algorithm: that eventual discrepancy between the estimated phase distribution and the actual one is only tributary to the noise at the pilots. Then, as we see, in suitable conditions (\eg small $L$) the one-shot \gls{cbc} algorithm may outperform more sophisticated algorithms such as \gls{skr}.

\item 
The one-shot CBC+EP algorithms outperform the \gls{skr} phase-tracking and again, the explanation should be sought in the different propensity to the phase-slip phenomenon, which we conjecture is due to the following:
\begin{itemize}
\item The \gls{skr} algorithm obtains $\alpha_n(\theta)$ (or $\beta_n(\theta)$) using only the past message $\alpha_{n-1}(\theta)$ see \eqref{z.alpha.check} (or future ones, $\beta_{n+1}(\theta)$, see \eqref{z.beta.check}), and \emph{does not use} any of the messages  $\beta_{n}(\theta)$ (or $\alpha_n(\theta)$). On the other hand, the CBC+\gls{ep} algorithm, to calculate $\gamma_n(\theta)$, relies \emph{both} on $\alpha^{(i-1)}(\theta)$ and on $\beta^{(i-1)}_n(\theta)$, see \eqref{equal.proba.approx.2}. Using for each $n$, the future and the past observations  increases the robustness of the CBC+EP algorithm.

\item In the \gls{skr} algorithm, any phase slip in $\alpha_{n}(\theta)$  will propagate to $\alpha_{n+1}(\theta)$ and  $\alpha_{n+2}(\theta)$, etc. (the same reasoning applies to the messages $\beta_n(\theta)$, $\beta_{n-1}(\theta)$, etc.). In contrast, the \gls{ep} algorithm finds the projections $\tilde\gamma^{(i)}_n(\theta)$ independently for all $n$. Of course, there is some dependence between the projections because they are based on the same estimates $\alpha^{(i-1)}_n(\theta)$ and $\beta^{(i-1)}_n(\theta)$ obtained in the previous iteration but it does not lead to the propagation of the errors as in the \gls{skr} algorithm. 
\end{itemize}

In fact, the propensity of the \gls{skr} algorithm to produce the phase-slip should be seen as a  reason behind the moderate improvements due to the \gls{ep}  as observed in the case of SKR+EP(2) in \secref{Sec:SKR.Numerical}.

\item In the hybrid algorithm CBC+SKR both, the past and the future are exploited at each $n$ and this removes the inefficiencies of the \gls{skr} algorithm we mentioned above: for $L=25$, the  CBC+SKR algorithm performs well both for high and low \gls{snr} behaving similarly as the best of CBC+EP(2) and CBC+EP(3). For $L=17$, the results of CBC+SKR algorithm are the same as those obtained by means of the algorithm CBC+EP(3). Note that the complexity of the algorithms CBC+EP(3) and CBC+SKR is also virtually the same.

\item The CBC+EP and CBC+SKR algorithms behave similarly for 16-\gls{qam} and 64-\gls{qam}. On the other hand, the poor performance of the \gls{skr} algorithm we observe for 16-\gls{qam}, disappears for 64-\gls{qam} transmission above $\PER=10^{-2}$ (when $L=25$) and above $\PER=10^{-3}$ ($L=17$).  The propensity of the \gls{skr} algorithm to produce the phase-slips manifests for $\SNRrv>21$dB.

Explaining formally for such a behaviour requires more investigation. Nevertheless, it should be noted that keeping the coding rate constant ($r=\frac{7}{8}$) and increasing the constellation size (from $M=16$ to $M=64)$ provides a larger margin between the theoretical value of the \gls{mi} required for decoding, \ie $rm$, and the maximum theoretical value, $m$, that the \gls{mi} can achieve in the transmission of the $M$-ary constellation. We may then hypothesize that larger \gls{mi} margin translates into smaller probability of decoding error (\eg caused by the phase slips) and thus the differences between the algorithms appear for low values of the \gls{per}.

%
%
\end{enumerate}

\subsection{Iterative receivers}

In the iterative algorithms we limit our considerations to a particular form of scheduling of operations between the demodulation/phase-tracking and the decoder: each demodulation/phase-tracking operation (where, the \gls{ep} phase tracking is iterative in itself) is followed by one decoding iteration.  

This is not necessarily the best solution but was extensively used in the literature, \eg \cite{Colavolpe05,Shayovitz16,Kreimer18} and thus constitues a well-accepted basis for comparison.  This means that the iterative receiver has to run the demodulation/phase tracking  $I_\tr{dec}$ times. The decoder's state (captured by the internal messages from the check to variables nodes) is preserved between iterations. 

As a reference  for the iterative receivers we also can use the curve ``All-pilots''. It is obtained assuming that, during the demodulation of the symbol $x_l$, all the symbols in $\set{x_l}_{l=1}^N$ but $x_n$ are pilots. This corresponds to a hypothetical situation when the output of the decoder provides highly reliable information about the bits which gives certainty about the symbols $\bx$ (that is, $P^{\tr{a}}_n(a)\approx 1$ for $a=x_n$); these, in turn, become de facto pilots allowing for the best possible phase tracking.  

Then, when calculating $\hat{P}_n(a)$, the only uncertainty about the symbol $x_n$ is caused by the \gls{awgn} as well as  by the residual phase noise -- this part of it which cannot be estimated from other symbols $x_l, l\neq n$. The ``All-pilots" curve is a performance limit for all joint phase-tracking and demodulation/decoding algorithms we study here\footnote{More precisely, for all those algorithms which do not use the decoder's feedback in the \gls{llr} calculation in \eqref{lambda.n.l}; see comments at the end of \secref{sec:model}.} and is the same irrespectively of the pilot spacing $L$. If the iterative \gls{dmp} is too complex to implement, the ``All-pilots" curve may be treated as its proxy. 

\begin{figure}[bt]
\begin{center}
\psfrag{DMP-T}[lc][lc][\siz]{DMP}
\psfrag{DMP10XXX.XXX.X}[lc][lc][\siz]{DMP one-shot}
\psfrag{CBC1S}[lc][lc][\siz]{CBC one-shot}
\psfrag{CBC+GA}[lc][lc][\siz]{CBC+GA \cite{Colavolpe05}}
\psfrag{CBC+CMM}[lc][lc][\siz]{CBC+CMM}
\psfrag{All-pilots}[lc][lc][\siz]{All-pilots}
\psfrag{CBC+CMM 0.7}[lc][lc][\siz]{CBC+CMM, $\psi=0.7$}
\psfrag{CBC+GA 0.7}[lc][lc][\siz]{CBC+GA, $\psi=0.7$}
\psfrag{ylabel}[lc][lc][\siz]{$\PER$}
\psfrag{xlabel}[lc][lc][\siz]{$\SNRrv$ [dB]}
\scalebox{\sizf}{\includegraphics[width=\sizfs\linewidth]{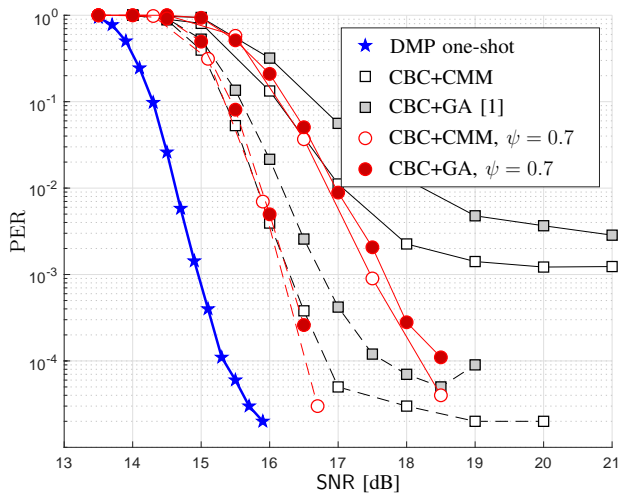}}
\end{center}
\caption{The iterative \gls{cbc} algorithms for $16$-\gls{qam} transmission and two different projection methods, the \acrfull{ga} and the \acrfull{cmm}. Both exhibit error floor which is removed when the \glspl{llr} are scaled down via \eqref{llr.scaling} with $\psi=0.7$; $L=17$ (dashed lines) and $L=25$ (solid lines).}\label{fig:PER.Iterative.CBC}
\end{figure}

We only consider an example of 16-\gls{qam} and  start with a short analysis of the results obtained using the canonical \gls{cbc} algorithm. They are shown in \figref{fig:PER.Iterative.CBC}, where two ways of carrying out the projection are considered: the \gls{ga} (results denoted by \gls{cbc}+\gls{ga}) and the \gls{cmm} (\gls{cbc}+\gls{cmm}).

While the  difference between the \gls{cbc}+\gls{ga} and the \gls{cbc}+\gls{cmm} algorithms is well notable (which indicates the importance of the approximations/projection we make in the phase tracking), the qualitative behaviour of the \gls{cbc} algorithms is the same and the gap to the \gls{dmp} remains large. 

More importantly, the error floor appears in the \gls{per} curve, a phenomenon which was already observed in \cite[Fig.~8]{Kreimer18}. In fact, with increasing \gls{snr} we observed even a slight deterioration of the \gls{per} curve. 

The error floor may be also observed for $L=17$; here we show the results for extended scale of the \gls{per} and they should be read with caution because we only simulate the transmission of $10^5$ blocks; despite somewhat  erratic behaviour of the \gls{per} curve, the error floor is visible.

Its origin should be sought  in unreliable phase approximations due to the parametric \gls{mp}. The decoding is then based on the \glspl{llr} calculated from  unreliable phase estimates and will produce erroneous estimates of the coded bits; this may be seen as overconfident decoding of a wrong codeword. 

In the iterative receivers this effect is amplified because the extrinsic \glspl{llr} of the decoder become prior \glspl{llr}, $\lambda^\tr{a}_{n,k}$ defined for all coded bits $c_{n,k}$. This phenomenon does not need to occur systematically in each transmission and is not equally detrimental in all phase-tracking algorithms. In fact, the frequency of its occurrence will be measured by the level of the error floor: $\PER\approx 10^{-3}$ for $L=25$ and $\PER\approx 10^{-5}$ for $L=17$;  this effect also occurs slightly more often when the \gls{ga} is used rather than the \gls{cmm}; this indicates that the approximation  $\tilde\gamma_n(\theta)$ obtained by means of the \gls{cmm} represents  $\gamma_n(\theta)$ more faithfully than does the \gls{ga}.

To deal with incorrectly calculated \glspl{llr}, a well-known approach relies on scaling them down\footnote{Akin to the scaling of the \glspl{llr} in the max-log decoders \cite{Xu14}.} as
\begin{align}\label{llr.scaling}
\tilde\lambda^\tr{a}_{n,k}=\psi \lambda^{a}_{n,k},
\end{align}
and next using them to calculate the prior symbol log-probabilities $\hat{P}^\tr{a}_n(a)$ via \eqref{Pa.lambda}.

As shown in \figref{fig:PER.Iterative.CBC}, this strategy efficiently removes the error floor. We used $\psi=0.7$ but very similar outcomes were obtained for $\psi\in(0.6,0.9)$; this indicates that the operational condition of the \gls{cbc} are set at the limit of reliable decoding; then, scaling down by any ``reasonably"  value $\psi$ offsets the negative effect of unreliable phase estimates and removes the error floor.

\begin{figure}[bt]
\begin{center}
\psfrag{DMP-T}[lc][lc][\siz]{DMP}
\psfrag{DMP10XXX.XX}[lc][lc][\siz]{DMP one-shot}
\psfrag{All-pilots}[lc][lc][\siz]{All-pilots}
\psfrag{Intra-MP}[lc][lc][\siz]{\gls{skr}}
\psfrag{HYBRID}[lc][lc][\siz]{CBC+SKR}
\psfrag{EP2}[lc][lc][\siz]{CBC+EP(2)}
\psfrag{EP3}[lc][lc][\siz]{CBC+EP(3)}
\psfrag{ylabel}[lc][lc][\siz]{$\PER$}
\psfrag{xlabel}[lc][lc][\siz]{$\SNRrv$ [dB]}
\scalebox{\sizf}{\includegraphics[width=\sizfs\linewidth]{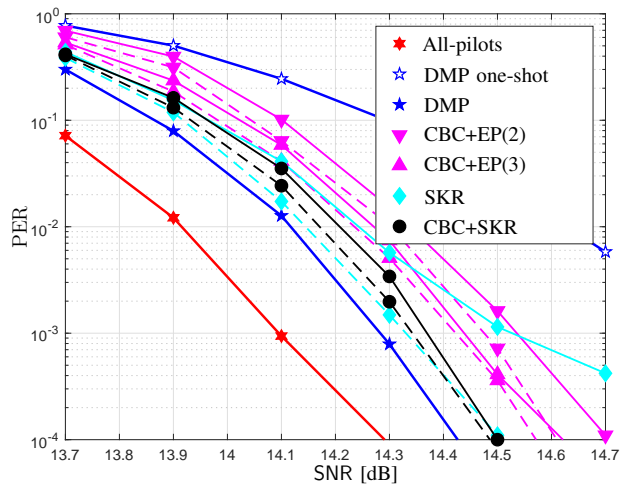}}
\end{center}
\caption{Performance of the iterative phase tracking algorithms in $16$-\gls{qam} transmission: CBC+EP($I_\tr{ep}$) (\secref{Sec:It.EP}) and SKR (\secref{Sec:Intra-MP}); two pilot spacings $L=17$ (dashed) and $L=25$ (solid), are considered; all algorithms (but \gls{dmp}) use \eqref{llr.scaling} with $\psi=0.7$. The  ``All pilots'' curve is the performance limit for any joint phase tracking and decoding.}\label{fig:PER.Iterative.all}
\end{figure}

The difference between the \gls{ga} and the \gls{cmm} results is very small after this modification yet well notable especially for $L=25$. More importantly, the results are still far from the performance of one-shot \gls{dmp} and thus we focus in \figref{fig:PER.Iterative.all} on the iterative phase tracking algorithms CBC+EP($I_\tr{ep}$), SKR, and CBC+SKR. The results obtained  are in line with those previously reported in the literature, \eg \cite{Shayovitz16,Kreimer18}: the performance is notably improved approaching the limits of the iterative \gls{dmp} where only the \gls{skr} algorithm shows the tendency to error floor for $L=25$. 

However, the differences in the results, being measured by fractions of dB are of  little practical importance thus, if we decide to use the decoder's feedback, all algorithms are practically equally suitable for implementation. The choice should be merely guided by the complexity which is the lowest for the CBC+EP(2) algorithm.

\section{Conclusions}\label{sec:Conclusions}
In this work we dealt with the problem of phase tracking in single-carrier transmission. This problem  was often solved in the literature via \acrfull{mp} on the graph which describes the relationship between all involved random variables;  the performance of the \acrfull{dmp} is then considered a reference for the performance.

The algorithms from the literature, inspired mostly by the \acrlong{cbc} algorithm \cite{Colavolpe05} rely heavily on the feedback from the decoder. This is also the case for the  \acrlong{skr} algorithm we defined here and which succinctly summarizes the works of \cite{Shayovitz16} and \cite{Kreimer18}.

With that regard our work is different as we focused on the signal processing approach to the phase-tracking that does not require decoder's feedback. This  approach is much more in line with the current industrial practice, which separates the operation of the modulation and decoding.

We proposed to improve the \gls{cbc} and \gls{skr} algorithms iteratively, via the \acrfull{ep}. Using numerical examples, we have shown that the algorithms we obtained outperform notably the \gls{cbc} and \gls{skr} algorithm before any feedback from the decoder is available. Moreover, the probabilistic framework allows us to design new algorithms via hybridization; and example  of such algorithms (dubbed CBC+SKR) is given and shown to perform particularly well.

Finally, we have also shown that the new algorithms result in efficient phase tracking when placed in the decoding loop.


\appendix
\subsection{Circular moment matching}\label{Sec:CMM}

The first circular moment of the circular distribution $g(\theta)$ is defined as
\begin{align}\label{M1}
\mu=\mfM[ g(\theta) ] 
&= \int_{0}^{2\pi}  g(\theta) \e^{\jj \theta} \dd\theta
\end{align}
and for the normalized Tikhonov distribution we obtain
\begin{align}\label{moment.Tikhonov}
\mfM\left[\frac{\mcT(\theta;z)}{2\pi I_0(|z|)}\right]&= B(|z|)\e^{\jj \angle z}=\frac{B(|z|)}{|z|}z,
\end{align}
where
\begin{align}
B(x) = \frac{I_1(x)}{I_0(x)},
\end{align}
$I_1(x)$ is the first-order Bessel function, and the last equation in \eqref{moment.Tikhonov} conveniently eliminates the need for calculation of the angle $\angle z$ (so non-linear functions are not needed).

Since $B(x)$ (or rather $B(x)/x$) will appear frequently and is not available in closed-form, we propose a simple approximation in \appref{Sec:Bx}.

\begin{definition}
For the  mixture of Tikhonov distributions
\begin{align}
g(\theta) = \sum_{a\in\mcA}  \e^{\eta(a)}\mcT(\theta;z(a)),
\end{align}
appearing \eg in \eqref{gamma.def.mixture}, \eqref{z.alpha.check}, \eqref{z.alpha.check}, \eqref{apost.theta.i1}, the \gls{cmm} projection on the space, $\mcF$, of Tikhonov distributions
\begin{align}
\mcT(\theta; z_{\tilde{g}})&=\mcF\big[\sum_{a\in\mcA}  \e^{\eta(a)}\mcT(\theta;z(a))\big],
\end{align}
is given by
\begin{align}
z_{\tilde{g}}&=\mf{CMM}\Big[ \set{\eta(a), z(a)}_{a\in\mcA}  \Big]\\
\label{z.tilde.def.CMM}
&=\frac{B^{-1}(|\mu_g|)}{|\mu_g|}\mu_g,
\end{align}
where
\begin{align}\label{mu.g.CMM.def}
\mu_g=\mfM\big[ g(\theta) \big]=\sum_{a\in\mcA} \xi(a) \frac{B(|z(a)|)}{|z(a)|}z(a)
\end{align}
and
\begin{align}
\label{xi.def.CMM}
\xi(a)&=\frac{\e^{\tilde\eta(a)-\tilde\eta_\tr{max}}}{\sum_{a\in\mcA} \e^{\tilde\eta(a)-\tilde\eta_\tr{max}}},\\
\label{tilde.eta.def.CMM}
\tilde\eta(a)&= \eta(a) + \hat{I}_0(|z(a)|),\\
\tilde\eta_\tr{max} &=\max_{a\in\mcA} \tilde\eta(a).
\end{align}
\end{definition}

To understand the derivation, we remind that the \gls{cmm} consists in finding a (normalized) Tikhonov distribution $\tilde{g}(\theta;z_{\tilde{g}})=\mcT(\theta;z_{\tilde{g}})/(2\pi I_0(|z_{\tilde{g}}|))$ closest, in the sense of the \gls{kl} distance, to a distribution  $g(\theta)$. This amounts to solving the  following optimization problem:
\begin{align}
z_{\tilde{g}}
&=\argmin_{z} \int_0^{2\pi} g(\theta)\log\frac{g(\theta)}{\tilde{g}(\theta;z)} \dd \theta\\
&=\argmax_{z}  \int_0^{2\pi} g(\theta) \Re[z\e^{-j\theta}] \dd \theta - \log I_0 (|z|)\\
&=\argmax_{z}  \Re[z\mu_g^*] - \log I_0 (|z|)\\
\label{Moment.matching}
&=B^{-1}(|\mu_g|)\e^{\jj \angle \mu_g }=\frac{B^{-1}(|\mu_g|)}{|\mu_g|}\mu_g,
\end{align}
where $B^{-1}(\cd)$ is the inverse of $B(\cd)$ and, again we preferred the angle-free formulation in \eqref{Moment.matching}.

Thus, \eqref{Moment.matching} explains \eqref{z.tilde.def.CMM}. 

We then need to calculate the circular moment, $\mu_g$, of $g(\theta)$ so the latter must be represented as a sum of normalized Tikhnonov distributions
\begin{align}
g(\theta)\propto\sum_{a\in\mcA} \e^{\tilde\eta(a)}\frac{\mcT(\theta;z(a))}{2\pi I_0(|z(a)|)},
\end{align}
where $\tilde\eta(a)$ is given by \eqref{tilde.eta.def.CMM}. And since $g(\theta)$ must be normalized, we require that
\begin{align}
g(\theta)=\sum_{a\in\mcA} \xi(a)\frac{\mcT(\theta;z(a))}{2\pi I_0(|z(a)|)},
\end{align}
where $\sum_{a\in\mcA}\xi(a)=1$, and this explains \eqref{xi.def.CMM} in which where we also introduced $\tilde\eta_\tr{max}$ to avoid overflows in numerical implementations.

Using simple algebra we find that, for a particular case of the large-argument approximation used in  of $B(\cd)$ and $B^{-1}(\cd)$,  see \eqref{B.approx} and \eqref{B.inv.approx}, the expressions \eqref{z.tilde.def.CMM} and \eqref{mu.g.CMM.def} are equivalent to those shown in \cite[Eq.~(100) and Eq.~(101)]{Shayovitz16}. 

On the other hand, our expressions, being explicitly based on $B(\cd)$ and $B^{-1}(\cd)$ are valid for any argument (not only large values) so using  them is much safer as it eliminates potential errors due to violation of approximation conditions. For example, it allowed us to find that the approximation \cite[Eq. (102)]{Shayovitz16} is the source of deteriorated performance, see \secref{Sec:SKR.Numerical}.

\subsection{Approximation of the function $B(x)=\frac{I_1(x)}{I_0(x)}$}\label{Sec:Bx}

\begin{figure}[tb]
\begin{center}
\psfrag{exact}[lc][lc][\siz]{$B(x)$}
\psfrag{approx}[lc][lc][\siz]{\eqref{B.approx}}
\psfrag{xlabel}[lc][lc][\siz]{$x$}
\scalebox{\sizf}{\includegraphics[width=\sizfs\linewidth]{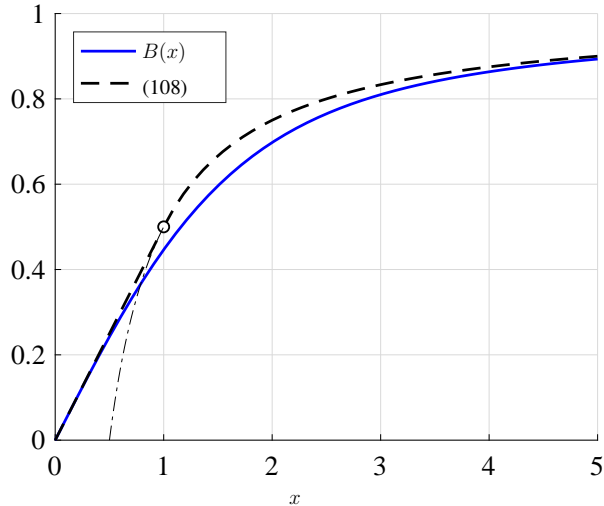}}
\end{center}
\caption{Comparison between $B(x)$ (solid) and its two-interval approximation from \eqref{B.approx} (dashed). The interval limit at which the two approximations merge is shown with a circle. The approximation based on the large-argument value is shown as well (dashed-dotted) but only for $x\ge \frac{1}{2}$ as the function becomes negative for $x<\frac{1}{2}$.}\label{fig:Bessel.B}
\end{figure}

We approximate $B(x)$ as
\begin{align}
\label{B.approx}
B(x)&\approx 
\begin{cases}
\frac{1}{2}x & \text{if} \quad x\leq 1\\
1-\frac{1}{2x}& \text{if}\quad  x>1
\end{cases}.
\end{align}
The large-argument approximation in \eqref{B.approx} (case $x>1$) is known, \eg \cite[Appendix~A]{Shayovitz16} but becomes negative for $x<0.5$. So, while most often $x$ is assumed large, \eg $x> 2$   \cite[Appendix~A]{Shayovitz16} -- in which case the large-argument approximation is quite precise, see \figref{fig:Bessel.B}, it is not uncommon to obtain $x\approx 0$. In such a case the large-argument approximation leads to approximation errors which are quite pernicious: even if approximating $B(x)$ with a negative number $B(x)\approx 1-\frac{1}{2x}$ makes no sense, such an error is difficult to spot in numerical implementation as it may appear in the sum of complex numbers: see, for example \eqref{mu.g.CMM.def} in the moment-matching procedure described in \appref{Sec:CMM}. 

In our simulation, such errors were relatively rare and did not lead to significant differences in performance, nevertheless, we believe it is much more sound to use the approximation which covers the entire range of the admissible arguments. This issue motivates us to introduce the small-argument approximation (for $x\leq1$) which is obtained via Taylor series development of $B(x)$ around $x=0$. Both, the large-, and the small-argument approximations meet at $x=1$ which is the threshold for using one approximation or another in \eqref{B.approx}. 

Of course, for the purpose of calculation of the circular moment, we rather need
\begin{align}
\label{B.approx.div}
\frac{B(x)}{x}&\approx 
\begin{cases}
\frac{1}{2} & \text{if} \quad x\leq 1\\
\frac{1}{x}(1-\frac{1}{2x})& \text{if}\quad  x>1
\end{cases}
\end{align}
and 
\begin{align}
\label{B.inv.approx}
\frac{B^{-1}(y)}{y}&\approx 
\begin{cases}
2 & \text{if} \quad y\leq \frac{1}{2}\\
\frac{1}{2y(1-y)}& \text{if}\quad  y>\frac{1}{2}
\end{cases}
\end{align}
obtained inverting \eqref{B.approx}.

We show in \figref{fig:Bessel.B} the function $B(x)$ and its approximation \eqref{B.approx} which gives us an idea about the approximation error. More importantly, and unlike the previously used large-argument approximation which was meaningful (\ie non-negative) only for $x>\frac{1}{2}$, we cover all range of the argument $x$ using simple functions. 

\subsection{Convolving Tikhonov and Gaussian distributions}\label{conv.T.G}

The convolution of the Tiknonov and the circular Gaussian distributions may be obtained by moment matching principle \cite[Lemma~3]{Kurz16}: the result of the convolution should have the same moment as the product of the moments of the convolved distributions, \ie
\begin{align}\label{conv.TT.MM}
\mcT(\theta;\tilde{z})&\approx\mcT(\theta;z)\convop\omega(\theta)\\
\mfM\left[\frac{\mcT(\theta;\tilde{z})}{2\pi I_0(|\tilde{z}|)}\right]&=\mfM\left[\frac{\mcT(\theta;z)}{2\pi I_0(|z|)}\right]\mfM\left[\omega(\theta)\right]
\end{align}

Since $\mfM[\omega(\theta)]=\e^{-\frac{\sigma_w^2}{2}}\approx(1-\frac{\sigma^2_w}{2})$, where  the approximation  is valid for small $\sigma^2_w$, using \eqref{moment.Tikhonov} we can  write
\begin{align}
B(|\tilde{z}|)\e^{\jj \angle \tilde{z}}&\approx B(|z|)\e^{\jj\angle z}\Big(1-\frac{\sigma^2_w}{2}\Big)\\
\label{conv.TT.z3.app}
\tilde{z}&\approx B^{-1}\left(B(|z|)\big(1-\frac{\sigma^2_w}{2}\big)\right)\e^{\jj \angle z}\\
\label{z.tilde.C}
&\approx
\begin{cases}
\frac{z}{|z|\sigma^2+1} & \text{if}\quad |z|>1\\
z(1-\frac{\sigma^2}{2}) & \text{if}\quad |z| \leq 1,
\end{cases}
\end{align}
where \eqref{z.tilde.C} is obtained using large- and small-argument approximations of $B(x)$, shown in \eqref{B.approx}.

We note that the large-argument approximation  ($|z|>1$) is known, see \cite{Colavolpe05}\cite{Barbieri07}\cite{Shayovitz16}, and using it did not affect (significantly) the results in our simulations but, again, rigorous approximations, valid for the entire range of input argument are, in general, more sound and useful.


\end{document}